\documentclass[MSNbibl,nameyear,dvips]{arxstspdf}
\usepackage{dcolumn}
\usepackage{subenv}
\usepackage{graphicx}
\usepackage{flushend}
\usepackage{stfloats}

\volume{27}
\issue{3}
\pubyear{2012}
\firstpage{319}
\lastpage{339}
\doi{10.1214/11-STS381} 

\makeatletter
\def\figurename{Display}
\newcolumntype{d}[1]{D{.}{.}{#1}}
\makeatother

\begin{document}
\begin{frontmatter}

\title{Multivariate Bayesian Logistic Regression
for Analysis of Clinical Study Safety Issues\thanksref{T1}}
\relateddois{T1}{Discussed in \relateddoi{d}{10.1214/12-STS381A},
\relateddoi{d}{10.1214/12-STS381B} and
\relateddoi{d}{10.1214/12-STS381C};
rejoinder at \relateddoi{r}{10.1214/12-STS381REJ}.}
\runtitle{MBLR for Clinical Safety Data}

\begin{aug}
\author[a]{\fnms{William} \snm{DuMouchel}\corref{}\ead[label=e1]{bill.dumouchel@oracle.com}}
\runauthor{W. DuMouchel}

\affiliation{Oracle Health Sciences}

\address[a]{William DuMouchel is Chief Statistical Scientist, Oracle Health
Sciences, Van De Graaff Drive, Burlington, Massachusetts 01803, USA
\printead{e1}.}

\end{aug}

\begin{abstract}
This paper describes a method for a model-based analysis of clinical
safety data called multivariate Bayesian logistic regression (MBLR).
Parallel logistic regression models are fit to a set of medically
related issues, or response variables, and MBLR allows information from
the different issues to ``borrow strength'' from each other. The method
is especially suited to sparse response data, as often occurs when
fine-grained adverse events are collected from subjects in studies sized
more for efficacy than for safety investigations. A combined analysis of
data from multiple studies can be performed and the method enables a
search for vulnerable subgroups based on the covariates in the
regression model. An example involving 10 medically related issues from
a pool of 8 studies is presented, as well as simulations showing
distributional properties of the method.
\end{abstract}


\end{frontmatter}

\section{Introduction}
\vspace*{4pt}

This paper introduces an analysis method for safe\-ty data from a pool of
clinical studies called multivariate Bayesian logistic regression
analysis (MBLR). The dependent or response variables in the MBLR are
defined at the subject level, that is, for each subject the response is
either 0 or 1 for each safety issue, depending on whether that subject
has been determined to be affected by that issue based on the data
available at the time of the analysis. Safety issues can include
occurrence of specific adverse events as well as clinically significant
lab tests or other safety-related measurements. The predictor variables,
assumed to be dichotomous or categorical, are all assumed to be
observable at the time of subject randomization. The analysis is
cross-sectional rather than longitudinal, and does not take into account
the variability, if any, of the length of time different subjects have
been observed. The primary predictor is study Arm, assumed to be
dichotomous with values ``Treatment'' or ``Comparator.'' Other
subject-level covariates may be included, such as gender or age
categories, or medical history variables. One feature of the MBLR
approach is that the interactions of treatment arm with each of the
other covariates are automatically included in the analysis model. Data
from a pool of multiple studies (having common treatment arm
definitions) may be included in the same analysis, in which case the
study identifier would be considered a subject covariate. Analyses
involving a pool of studies are similar in spirit to a~full-data
meta-analysis.

Estimation of effects involves a hierarchical Bayesi\-an algorithm as
described below. There are two primary rationales for the Bayesian
approach. First, data concerning safety issues are often sparse, leading
to high variability in relative rates of rare events among subject
subgroups, and the smoothing inherent in empirical Bayes shrinkage
estimates can alleviate problems with estimation of ratios of small
rates and the use of multiple post-hoc comparisons when encountering
unexpected effects. Second,\break MBLR fits the same analytical model to each
response variable and then allows the estimates of effects for the
different responses to ``borrow strength'' from each other, to the
extent that the patterns of coefficient estimates across different
responses are similar. This implies that the different safety issues
should be medically related, so that it is plausible that the different
issues have related mechanisms of causation or are different expressions
of a broad syndrome, such as being involved in the same body system, or
different MedDRA terms in nearby locations in the MedDRA hierarchy of
adverse event definitions. The goal is to assist with the problem of
uncertain \textit{granularity of analysis}. The question of how to
classify and group adverse drug reaction reports can be controversial
because different assignments can change the statistical significance of
count data treatment effects, and methods and definitions for comparing
adverse drug event rates are not well standardized (Dean, \citeyear{r3}).
Sometimes the amount of data available for each of the related safety
issues is too little for reliable comparisons, whereas doing a~single
analysis on a transformed response, defined as present when any of the
original issues are present, risks submerging a few potentially
significant issues among others having no treatment association. The
Bayesian approach is a compromise between these two extremes.

The proposed methodology is not intended to replace or replicate other
processes for evaluating safe\-ty risk but rather to support and augment
them. In spite of the formal modeling structure, its spirit is more a
mixture of exploratory and confirmatory analysis, a way to get a big
picture review when there are very many parameters of interest. The
resulting estimates with confidence intervals can provide a new approach
to the problem of how to best evaluate safety risk from clinical studies
designed to test efficacy.

This paper describes the statistical model and the estimation algorithms
used in a commercial implementation of MBLR. There is also some
discussion of alternate models and algorithms with reasons for our
choices. An example analysis utilizes data from a set of clinical
studies generously provided by an industry partner, and a simulation
provides information on the statistical properties of the method.

\section{The Bayesian Model for MBLR}

As with standard logistic regression, MBLR produces parameter estimates
interpretable as log odds, and provides upper and lower confidence
bounds for these estimates. The method is based on the hierarchical
Bayesian model described below. Identical regression models (i.e., the
same predictor variables for different response variables) are estimated
assuming that the relationships being examined are all based on the same
underlying process. The response variables represent issues comprising a
potentially common safety problem and the underlying process is an
adverse reaction caused by the treatment compound. The regression models
are various examinations of relationships between subgroups defined by
the covariates and the response issues. The Bayesian estimates of
treatment-by-covariate interactions are conservative (estimates are
``shrunk'' toward null hypothesis values), in order to reduce the false
alarm due to high variance in small sample sizes. This conservatism is a
form of adjustment for multiple comparisons.

It is natural to desire a comparison of MBLR with a more standard
analysis, which, for the present purpose, means a logistic regression
model where the estimates for the different responses are not shrunk
toward each other, and where interactions between treatment and other
covariates are not being estimated. However, it can often happen, with
sparse safety data involving rare adverse events and the use of other
predictors in addition to the treatment effect, that standard logistic
regression estimation can fail, because the likelihood function has no
unique finite maximizing set of parameters. Gelman et al. (\citeyear{r4}) discuss
this problem, caused by what they call \textit{separation} and
\textit{sparsity}, and suggest the automatic use of weakly informative
prior distributions as a default choice for such analyses. Along the
lines of the Gelman et al. (\citeyear{r4}) suggestion, we will compare MBLR to a
``weak Bayes'' method that corresponds to setting certain variance
components (that are estimated by MBLR) to values selected to be so
large that the resulting estimates would be virtually the same as those
of standard logistic regression if the data\vadjust{\goodbreak} are not so sparse as to be
unidentifiable. This comparison method will be denoted
\textit{regularized logistic regression} (RLR).

The event data can be considered as a $K$-column matrix $Y$, with a row for
each subject and a column for each issue, and where $Y_{sk} = 1$ if
subject~$s$ experienced issue $k$, $0$ otherwise. Since all subject covariates
are assumed categorical, we will use a~grouped-data approach, where
there are $n_{i}$ subjects ($i =\break 1, \ldots, m$) that have identical
covariates and treatment allocation in the $i$th group, and where
$N_{ik}$ of these subjects experienced issue $k$.

MBLR requires the inclusion of the treatment arm and of one or more
additional predictors in the model, where all predictors are
categorical.

Across the set of issues a single regression model is used. If there are
$J$ predictors excluding Treatment, and the $j$th predictor has $g_{j}$
categories, $j = 1, \ldots, J$, then there will be $G = \sum g_{j}$
subgroups analyzed. The model will usually have $2G + 2 - 2J$ degrees of
freedom and estimation is performed by constraining sums of coefficients
involving the same covariate to add to 0. The Bayesian methods presented
here allow estimation in the presence of additional collinearity of
predictors, in which case the computed posterior standard deviations
would then reflect the uncertainty inherent in a deficient design.

For the $i$th group of subjects, the modeled probability of experiencing
issue $k$ is $P_{ik}$, where
\begin{eqnarray}
\label{eq1}P_{ik} &=& 1/[1 + \exp(- Z_{ik})]\quad\mbox{and where}
\\
\label{eq2}Z_{ik} &=& \alpha_{0k} +
\sum_{1\leq g\leq G} X_{ig} \alpha_{gk}\nonumber
\\[-8pt]
\\[-8pt]
&&{} +
T_{i} \biggl(\beta_{0k} + \sum_{1\leq g\leq G}
X_{ig} \beta_{gk}\biggr).\nonumber
\end{eqnarray}

The $G$ columns of $X$ define the $G$ dummy variables for the $J$ covariates,
and $T_{i}$ is an indicator for the treatment status of the $i$th group.
The values of $\alpha_{gk}$ ($g = 0, \ldots, G$; $k = 1, \ldots, K$)
define the risk of issue $k$ for the comparator subjects. As mentioned
above, the sums $\sum_{g} \alpha_{gk} = 0$, where the sums
are over the categories of each covariate for each $k$. The more natural
quantities ($\alpha_{0k} + \alpha_{gk}$) are the log odds
that a comparator subject in subgroup $g$ will experience issue $k$, $g = 1,
\ldots, G$, averaged across the categories of other predictors not
defined by subgroup~$g$.

Concerning treatment effects, the quantities ($\beta_{0k} + \beta_{gk}$)
 are the estimated log odds ratios for the risk of issue $k$
(treatment versus comparator) that subjects in group $g$ experience, $g =
1, \ldots, G$, averaged across the categories of other predictors not
defined by subgroup~$g$. The sums $\sum_{g} \beta_{gk}$ are
constrained just as the $\alpha$'s were.

If $G$ is large, there will be many possible subgroup comparisons, and,
since these confidence intervals have not been adjusted for multiple
comparisons, caution is advised in interpreting the largest few of such
observed subgroup estimates. The MBLR estimates of these quantities are
designed to be more reliable in the presence of multiple comparisons
because subgroup-by-treatment interaction effects are ``shrunk'' toward
0 in a statistically appropriate way, and there is also a partial
averaging across issues, so that subgroup and treatment effects and
subgroup-by-treatment interactions can ``borrow strength'' if there is
an observed similar pattern of treatment and subgroup effects in most of
the $K$ issues being analyzed. When configuring a multivariate Bayesian
logistic regression, the analyst should try to select those issues for
which there is some suspicion of a~common medical mechanism involved. If
the Bayes\-ian algorithm does not detect a common pattern of subgroup
effects, then the Bayesian algorithm will perform little partial
averaging across issues, because corresponding variance component
estimates will be large.

The Bayesian model is a two-stage hierarchical prior specification:
\begin{eqnarray}
\label{eq3}\alpha_{gk}|A_{g} &\sim& N(A_{g},\sigma_{A}^{2})
,\nonumber
\\[-8pt]
\\[-8pt]
 \eqntext{k = 1, \ldots, K ; g = 1, \ldots , G ,}
\\
\label{eq4}\beta_{0k}|B_{0} &\sim& N(B_{0}, \sigma_{0}^{2}) , \quad k = 1,
\ldots, K,
\\
\label{eq5}\beta_{gk}|B_{g} &\sim& N(B_{g}, \sigma_{B}^{2}) ,\nonumber
\\[-8pt]
\\[-8pt]
\eqntext{k = 1, \ldots, K ; g = 1, \ldots , G,}
\\
\label{eq6}B_{g} &\sim& N(0, \tau^{2}) ,\quad g = 1, \ldots ,G.
\end{eqnarray}

The prior distributions of $\alpha_{0k}$, $k = 1, \ldots, K$, and of
$A_{g}$, $g = 1, \ldots, G$, and of $B_{0}$ are assumed uniform within
($-\infty, +\infty$). Equations (\ref{eq3})--(\ref{eq5})
embody the assumption that
coefficients for the same predictor across multiple issues cluster
around the predictor-specific values ($A_{1}, \ldots, A_{G}, B_{0},
\ldots, B_{G}$) with the degree of clustering dependent on the
magnitude of three variances ($\sigma_{A}^{2}, \sigma_{0}^{2},
\sigma_{B}^{2}$). If any of these variances are near 0, there will
be a tight cluster of the corresponding regression coefficients across
the $K$ responses, whereas if they are large, there may be no noticeable
common pattern across $k$ for predictor $g$. The values of
$\alpha_{0k}$
correspond to the constant terms in\vadjust{\goodbreak} the regressions, and we assume no
common shrinkage of constant terms across issues, since the absolute
frequencies of the issues are not being modeled here.

Equation (\ref{eq6})
embodies the assumption that the null hypotheses $B_{g}
= 0$ (i.e., no treatment-by-covari\-ate interactions when averaged across
responses) are given priority in the analyses. This is the assumption
that helps protect against the multiple comparisons fallacy when
searching for vulnerable covariate subgroups. The value of $\tau^{2}$
determines how strongly to shrink the $G$ prior means $B_{g}$ toward 0
in the second level of the prior specification.

The four standard deviations ($\sigma_{A}, \sigma_{0},
\sigma_{B}, \tau)$ have prior distributions assumed to be uniform in
the four-dimensional cube $0 \leq \sigma$, $\tau \leq
d$. Their joint posterior distribution is approximated by a discrete
distribution for computational convenience, as described below. The
posterior distribution of the coefficients $\{A_{g}, B_{g},
\alpha_{gk}, \beta_{gk}\}$ is defined as a mixture of the
distributions of the coefficients conditional on the possible values of
the variance components. The method produces an approximate
variance--covariance matrix for all the coefficients, and this also
allows the estimation of standard deviations and confidence intervals
(credible intervals) for linear combinations of parameters such as the
quantities ($\beta_{0k} + \beta_{gk}$) describing the total
treatment effect estimates for each subgroup.

General discussion of hierarchical Bayesian regression models is
available in Carlin and Louis (\citeyear{r2}), although the particular model
(involving multiple responses) and estimation methods used in this paper
are not discussed there. Searle, Casella and McCulloch [(\citeyear{r8}), Chapter
9]
also discuss related methods, including a logit-normal model somewhat
similar to this one.

\section{Estimation Details}

\subsection*{Estimation of MBLR Parameters}

The estimation algorithm for MBLR is based on separate maximizations of
the posterior distributions of the coefficients, conditional on the
values of the variance components. Then these posterior distributions
are averaged to provide an overall posterior distribution, where the
weights in the average are determined by the Bayes factors for different
values of the vector of the four variance components. First we assume
that the four standard deviations ($\sigma_{A}, \sigma_{0},
\sigma_{B}, \tau$) are fixed and known and consider estimation of
the other parameters.

\subsection*{Estimation of Coefficients and Prior Means Conditional on Prior Standard
Deviations}

There are $M = 2(G+1)(K+1) - 1$ such parameters: $2G+1$ prior means,
$(G+1)K$
values $\alpha_{gk}$ and $(G+1)K$ values of $\beta_{gk}$. However,
$2J(K+1)$ sums of these parameters are defined as 0, leaving $M^* =
2(G-J+1)(K+1) - 1$ dimensions for estimation. It is convenient to
imagine that subjects are grouped according to unique values of their
covariates and treatment allocation, so that the data are the sample
sizes $n_{i}$ and the counts $N_{ik}$ ($i = 1, \ldots, m$; $k = 1,
\ldots, K$), where $i$ indexes $m$ strata defined by unique values of
covariates and treatment. The joint distribution of the parameters and
the data can be represented as
\begin{eqnarray}
&&p(A_{1},\ldots,A_{G}, B_{0}, \ldots,B_{G})\nonumber
\\
&&\quad {}\cdot\prod_{k}p(\alpha_{0k},\ldots,\alpha_{Gk},
\beta_{0k},\ldots, \beta_{Gk}|\{A\}\{B\})
\\
&&\hphantom{\quad {}\cdot\prod_{k}}{}\cdot p(\{N_{ik}\}|\{A\}\{B\}\{\alpha\}\{\beta\}). \nonumber
\end{eqnarray}

The prior distributions of $A_{1}, \ldots, A_{G}$, $B_{0}$ and the
$\{\alpha_{0k}\}$ are assumed uniform over ($-\infty, +\infty$),
whereas all the remaining parameters have prior distributions as given
in equations (\ref{eq3})--(\ref{eq6}).

Therefore, if $\log L$ is the log posterior joint distribution of all the
parameters, then, up to a constant,
\begin{eqnarray}\label{eq8}
2 \log L &=& - \biggl[\sum_{g>0}B_{g}^{2}/\tau^{2} + (G - J)
\log(\tau^{2})\biggr]\nonumber
\\
&&{}- \biggl[\sum_{g>0}\sum_{k}(\alpha_{gk} -
A_{g})^{2}/\sigma_{A}^{2}\nonumber
\\
&&\hspace*{25pt}{}+ (G - J)K
\log(\sigma_{A}^{2})\biggr]\nonumber
\\
&&{}- \biggl[\sum_{k}(\beta_{0k} - B_{0})^{2}/\sigma_{0}^{2} + K
\log(\sigma_{0}^{2})\biggr]\nonumber
\\[-8pt]
\\[-8pt]
&&{}- \biggl[\sum_{g>0}\sum_{k}(\beta_{gk} -
B_{g})^{2}/\sigma_{B}^{2}\nonumber
\\
&&\hspace*{24pt}{} + (G - J)K
\log(\sigma_{B}^{2})\biggr]\nonumber
\\
&&{}+ 2 \sum_{i}\sum_{k} [N_{ik} \log(P_{ik})\nonumber
\\
&&\hphantom{{}+ 2 \sum_{i}\sum_{k} [}{}+
(n_{i} - N_{ik})\log(1 - P_{ik})]. \nonumber
\end{eqnarray}

In (\ref{eq8}), the terms involving $\log(\tau^{2})$,
$\log(\sigma_{A}^{2})$ and $\log(\sigma_{B}^{2})$ all have a factor
($G - J$), rather than the more natural $G$, since there are $G$ values
$\{A_{g}\}$ and $\{B_{g}\}$. But since they are being estimated
subject to $J$ constraints where subsets of them add to $0$, the factor ($G
- J$) is substituted, analogous to the way REML estimates are defined
for variance components in a frequentist analysis. For fixed variance
components, maximization of (\ref{eq8}) with respect to all other parameters,
remembering that the $P_{ik}$ are defined by (\ref{eq1}) and
(\ref{eq2}), involves a
relatively straightforward modification of the usual logistic regression
calculations. The prior means $\{A_{g}\}$ and $\{B_{g}\}$ are
treated analogously to the coefficients $\{\alpha_{gk}\}$ and
$\{\beta_{gk}\}$ during the Newton--Raphson maximization of $\log L$.
Each iteration involves calculation of the vector $S$ of $M$ first
derivatives of $\log L$ with respect to the parameters in (\ref{eq8}) and the
$M\times M$ Hessian matrix $H$ of the negative second derivatives of $\log L$.
The initial values of $\alpha_{0k}$ are $\log(N_{+k}/(n_{+} -
N_{+k}))$, $k = 1, \ldots, K$ (the subscript ``$+$'' means sum over the
values of $i$), whereas the initial values of all other parameters are 0.

Upon convergence of the maximization, the vari\-ance--covariance matrix of
the estimated parameters is assumed to be
\begin{equation}\label{eq9}
V = V(\sigma_{A}, \sigma_{0}, \sigma_{B}, \tau) = H^{-1}.
\end{equation}

[Actually, the matrix $H$ will be singular because of the constraints that
reduce the rank of $H$. The interpretation of (\ref{eq9}) is as follows. Define a
subset $\theta^*$  of $M^*$ parameters out of the $M$-vector $\theta$, where
one parameter from each constrained subset has been omitted, but will be
constrained to be equal to the negative of the sum of the other
parameters in its subset. Define the $M\times M^*$ matrix $Z$ that converts
from $\theta^*$ to $\theta$, that is, $\theta = Z \theta^*$. Then (\ref{eq9}) is
interpreted as $V = Z(Z^{t}HZ)^{-1}Z^{t}$. The same
transformation is used during the Newton--Raphson maximization of $\log L$.
Also, in (\ref{eq11b}) and later, the determinant of $V$ is computed as the
determinant of $V^* = (Z^{t}HZ)^{-1}$.]

The computation of $V$ as $H^{-1}$ uses the assumption that the counts
$\{N_{ik}\}$ are independent across both $i$ and $k$, conditional on the
parameters. The occurrence of different events in the same subject may
be connected via the parameters, but not otherwise correlated in this
model. If this assumption is violated, the variances in $V$ may be
underestimated. Since the $M$ parameters include both all the coefficients
as well as their prior means, the variances in $V$ for any one component
automatically include uncertainty due to correlation with all other
components. In particular, uncertainty in the prior means $\{A_{g},
B_{0}, B_{g}\}$ is taken account of in the estimated posterior
variances of the $\{\alpha_{gk}, \beta_{gk}\}$\vadjust{\goodbreak} (up to the
accuracy of the approximate multivariate normality of the joint
posterior distribution of the parameters).\looseness=1

\subsection*{Accounting for Uncertainty in the Prior Standard
Deviations}

The prior distribution of the set of possible values of
($\sigma_{A}, \sigma_{0}, \sigma_{B}, \tau$) is assumed to be
uniform within the four-dimensional cube with limits ($0, d$), where a
default value of $d = 1.5$ is selected as discussed below. A discrete
search method approximates the posterior distribution within this cube.
Before discussing the details, consider the situation where the prior
standard deviation vector $\phi = (\sigma_{A}, \sigma_{0},
\sigma_{B}, \tau)$ is assumed to be one of $S$ discrete values
$\phi_{1}, \phi_{2}, \ldots, \phi_{S}$. Denote the vector of
coefficients and prior means by $\theta = (A_{1}, \ldots, A_{G},
B_{0}, \ldots, B_{G}, \alpha_{01}, \ldots, \alpha_{GK},
\beta_{01}, \ldots,\break \beta_{GK})$, and assume that the maximized $\log
L$
and the estimated posterior mean and covariance matrix of~$\theta$ are
$(\log L_{s}, \theta_{s}, V_{s})$ if $\phi = \phi_{s}$, $s
= 1, \ldots, S$. Then the marginal posterior distribution of $\theta$,
adjusting for uncertainty in $\phi$, is assumed to be multivariate
normal with mean $\hat\theta$  and covariance matrix $V$, where%
\begin{subequation}\label{eq10}
\begin{eqnarray}\label{eq10a}
\hat\theta&=& \sum_{s} \pi_{s} \theta_{s},
\\
\label{eq10b}
V &=& \sum_{s} \pi_{s} [V_{s} + (\theta_{s} -\hat\theta
)(\theta_{s} - \hat\theta)^{t}],
\end{eqnarray}
\end{subequation}
and where $\pi_{s}$, the posterior weight given to $\phi =
\phi_{s}$, $s = 1, \ldots, S$, is defined by
\begin{subequation}\label{eq11}
\begin{eqnarray}\label{eq11a}
\pi_{s} &=& \mathit{BF}_{s} / (\mathit{BF}_{1} + \cdots + \mathit{BF}_{S}),
\\
\label{eq11b}
\mathit{BF}_{s} &=& \exp(\log L_{s}) \sqrt{\operatorname{det}(V_{s})}.
\end{eqnarray}
\end{subequation}

The quantity $\mathit{BF}_{s}$ is the (relative) \textit{Bayes factor} for the
hypothesis $\phi = \phi_{s}$. The usual definition of the Bayes
factor requires the integration of the joint likelihood over the space
of all parameters not specified by the hypothesis---in this case the
space of all~$\theta$.~Using the approximation of this likelihood as
proportional to a multivariate normal density with co\-variance matrix
$V_{s}$, and the known fact that volume under the multivariate
exponential form\break $\exp[-\theta^{t}(V_{s})^{-1}\theta /2]$ is
proportional to the square root of the determinant of $V_{s}$, the
definition of $\mathit{BF}_{s}$ is as given in (\ref{eq11}).
The approximation (\ref{eq11}) is
the standard Laplace approximation often used for numerical integration
in Bayesian methods. However, a different justification for computing
(\ref{eq11b}) in order to obtain estimates for variance components is given by
the theory of h-likelihood (Lee and Nelder, \citeyear{r5}; Lee, Nelder and
Pawitan,
\citeyear{r6}; Meng, \citeyear{r7}).

\subsection*{Selection and computation of the values ($\phi_{s}, \pi_{s}$),
$s = 1, \ldots, S$}

Representing the 4-dimensional naturally continuous distribution of
$\phi$ by a set of discrete points is a~challenge. Assuming a range of $d
= 1.5$ for each element of $\phi$ and a spacing of 0.1 would mean a~grid
of~$S = 15^{4} > 50{,}000$ points, the vast majority of\break which~would have
values of $\pi_{s}$ nearly 0. Determination of a set of just $S = 33$
points to represent the approximate posterior distribution of $\phi$ is
performed as outlined next. A logistic transformation is used to convert
the bounded cube $(0, d)^{4}$ to the unbounded region where all four
elements can range from ($-\infty, +\infty$) by defining
\begin{eqnarray}\label{eq12}\qquad
\lambda &=& (\lambda_{A}, \lambda_{0}, \lambda_{B},
\lambda_{\tau} )\quad \mbox{where}\nonumber
\\
\sigma_{A} &=& d/(1 + e^{-\lambda_A}),\quad \sigma_{0} = d/(1 +
e^{-\lambda_0}),
\\
 \sigma_{B} &=& d/(1 + e^{-\lambda_B}), \quad \tau =
d/(1 + e^{-\lambda_{\tau}} ). \nonumber
\end{eqnarray}

With this transformation, a uniform prior distribution on ($0, d$) for
each $\sigma$ corresponds to a prior distribution for each $\lambda$
over the real line of $f(\lambda) \propto \sigma(\lambda) (d -
\sigma(\lambda))$. The purpose of this transform is to allow simpler
search procedures that don't have to worry about boundary constraints,
as well as to make approximation of the posterior by a multivariate
normal distribution more accurate. Then the posterior density of
$\lambda$ is assumed to be
\begin{eqnarray}\label{eq13}
g(\lambda)& =& g(\lambda_{A}, \lambda_{0}, \lambda_{B},
\lambda_{\tau} )\nonumber\\
 &\propto&
f(\lambda_{A})f(\lambda_{0})f(\lambda_{B})f(\lambda_{\tau}
)
\\
&&{}\cdot \exp(\log L_{s}) \sqrt{\operatorname{det}(V_{s})},\nonumber
\end{eqnarray}
where $\log L$ and $V$ in (\ref{eq13}) are now functions of $\lambda$, and the
$\lambda$'s vary over ($-\infty, +\infty$).

The determination of the discrete distribution ($\phi_{s},\allowbreak
\pi_{s})$, $s = 1, \ldots, S$, is a five-step process:

\textit{Step} 1: Use the method of steepest ascent to find the value
$\lambda^{\mathrm{max}}$ that maximizes $g(\lambda)$ in (13). Derivatives of
g are computed numerically as first difference ratios with respect to
each of the four arguments. The starting value for the search is
$\lambda = (0, 0,\break 0, 0)$.

\textit{Step} 2: Construct a design of $S = 33$ $\lambda$-values by
adding 16 points on the surface of each of two concentric spheres
centered at $\lambda^{\mathrm{max}}$. The points on the inner sphere consist
of 8 star points, where one component of $\lambda$ is
$\lambda^{\mathrm{max}} \pm 2\delta_{0}$ and the other three components
equal $\lambda^{\mathrm{max}}$, and 8 half-fractional factorial points, where
all components are $\lambda^{\mathrm{max}}\pm \delta_{0}$. The points
on\vadjust{\goodbreak}
the outer sphere are similar to those on the inner sphere, except that
$\delta_{0}$ is replaced by $1.5\delta_{0}$ and the fractional factorial
points are from the opposite half fraction as the fractional factorial
points on the inner sphere. The default value of $\delta_{0} = 0.3$ on
the scale of $\lambda$. Visualizing the geometry of the design, if a
4-dimensional sphere has radius 1.5 times another, it encloses about 5
times the volume.

\textit{Step} 3: The double central composite design of Step~2 is
centered but not scaled to the actual distribution $g(\lambda)$. To find
the appropriate scale factors in each dimension, $\delta = (\delta_{1},
\delta_{2}, \delta_{3}, \delta_{4})$, for a better fitting design, a
quadratic response surface model is fit to values of $\log
g(\lambda)$ across the $S$ points of this initial design. The fitted
model is
\begin{equation}
\log g(\lambda) = c_{0} + \sum_{i}
c_{i}\lambda_{i} + \sum_{i\leq j}
c_{ij}\lambda_{i}\lambda_{j}.
\end{equation}

Now if the quadratic model fit exactly (i.e., if $g$ were exactly
multivariate normal), then the second-order coefficients $c_{ij}$
would specify the elements of the inverse of the posterior covariance
matrix of~$\lambda$. Accordingly, we get what are hoped to be
approximate posterior standard deviations by setting $\delta = \mathrm{vector}$
of square roots of the diagonal of $H^{-1}$, where
\begin{equation}\label{eq15}
2H = \left[\matrix{ 2c_{11} & c_{12} & c_{13} & c_{14}
\cr
 c_{12} &
2c_{22} & c_{23} & c_{24}
\cr
 c_{13} & c_{23} & 2c_{33} & c_{34}
 \cr
c_{14} & c_{24} & c_{34} & 2c_{44}}\right].
\end{equation}

\textit{Step} 4: Next a new design like that of Step~2 is constructed
except that the $\delta_{0}$ used in Step~2 for all 4 dimensions is
replaced by $\delta = (\delta_{1}, \delta_{2}, \delta_{3}, \delta_{4})$
from Step~3, so that the spheres are scaled differently in each
dimension. The values of $\log g(\lambda)$ are computed for
these 32 new points and a new quadratic response surface is fit to this
33-point final design. Let the peak of this fitted surface be denoted
$\lambda^{\mathrm{fit}}$, which will not exactly equal $\lambda^{\mathrm{max}}$,
and redefine $\delta = (\delta_{1}, \delta_{2}, \delta_{3}, \delta_{4})$
by using the coefficients from the new quadratic response surface in
(\ref{eq15}).

\textit{Step} 5: The discrete distribution defined by
$\{\lambda^{(s)},\allowbreak
g(\lambda^{(s)}), s = 1, \ldots, S\}$ as computed in Step~4 will
rough\-ly approximate the continuous distribution defined by $g(\lambda)$,
but the approximation can be improved by modifying the $S = 33$
probabilities to constrain the 4 means and 4 standard deviations of the
discrete distribution to exactly match the values
$\lambda^{\mathrm{fit}}$
and $\delta$ that were computed from the response surface fit of Step~4.
The final probabilities $\pi_{s}$, $s = 1, \ldots, S$, are computed
as the solution to the following constrained optimization problem:

Find positive $\pi_{1}, \ldots, \pi_{S}$ that minimize the
Kul\-lback--Leibler divergence
\begin{eqnarray}\label{eq16}\qquad
&&\mathit{KL} = \sum_{s} g\bigl(\lambda^{(s)}\bigr)
\log\bigl[g\bigl(\lambda^{(s)}\bigr)/\pi_{s}\bigr],\nonumber
\\
&&\quad\mbox{subject to the 9-dimensional
constraints}\nonumber
\\[-8pt]
\\[-8pt]
&&\sum_{s} \pi_{s} = 1;\quad \sum_{s}
\pi_{s}\lambda^{(s)} = \lambda^{\mathrm{fit}};\nonumber
\\
&&\sum_{s}
\pi_{s}\bigl(\lambda^{(s)} - \lambda^{\mathrm{fit}}\bigr)^{2} =
\delta^{2},\nonumber
\end{eqnarray}
where the last two equations are each interpreted as 4 constraints, one
for each component of $\lambda$. The constrained minimization problem of
(\ref{eq16}) is solved using the method of Lagrange multipliers combined with a
Newton--Raphson solution of the resulting 9 equations.

Thus, the $\{\pi_{s}\}$ used in~(\ref{eq10}) are the solution to~(\ref{eq16}) rather
than the more direct values in~(\ref{eq11}). They differ from~(\ref{eq11}) by
incorporating the Jacobian terms of~(\ref{eq13}) and the further modifications
needed to satisfy the constraints in~(\ref{eq16}). The values of
$\{\phi_{s}\}$ used in~(\ref{eq10}) are the back-transformations defined
by~(\ref{eq12}) of the final $S = 33$ points $\{\lambda_{s}\}$ used in Steps~4~and~5.

\subsection*{Estimates Using Regularized Logistic Regression (RLR)}

To compare the MBLR results to standard logistic regression, and still
be able to avoid problems with nonidentifiability, as discussed above,
the RLR algorithm is defined by fitting MBLR under the constraints
\begin{equation}\label{eq17}
\hspace*{21.99pt}\sigma_{A} = 5,\quad \sigma_{0} = 5,\quad \sigma_{B} = 0.001,\quad \tau =
0.001.
\end{equation}

Setting $\sigma_{B}$ and $\tau$ very close to $0$ effectively
constrains the estimates of covariate-by-treatment inter\-actions to be 0.
Setting $\sigma_{A}$ and $\sigma_{0}$ to be very large~pre\-vents the
estimates across different response events from shrinking toward each
other The rationale for thinking that a prior standard deviation of 5 is
very large for a logistic regression coefficient is as follows.
Remembering that the coefficients are interpreted as logs of odds
ratios, an increase of 5 in a~coefficient corresponds to a
multiplicative factor of $e^{5} = 148.4$ in an odds ratio. With respect
to the assumed normal prior distributions in equations
(\ref{eq3})--(\ref{eq6}), the prior
standard deviation of 5 implies that about one-third of all estimated
odds ratios are expected to be outside the range of ($1/148=0.007$, 148).
This certainly seems to be well beyond the range of expected odds ratios
in any medical risk estimation situation. See Gelman et al. (\citeyear{r4}) for
a~related discussion. [In the Bayesian setup described above, we use as
default limits for the prior standard deviations ($0$, $d = 1.5$).
Considering a prior standard deviation to be as large as 1.5, where
$e^{1.5} = 4.5$, implies that about one-third of the estimated odds
ratios would be outside the range of ($1/4.5=0.22$, 4.5), which seems a
bit of a stretch, but barely conceivable.]

Using the values in (\ref{eq17}) for the prior standard deviations, this
alternative weak Bayesian prior method estimates the parameters and
their variances using the iterative Newton--Raphson estimation described
above. The resulting estimates are computationally reliable even if many
of the response events are sparse. Such estimates perform very little
shrinkage across response models because the prior standard deviations
in equations (\ref{eq3})--(\ref{eq6}) are large compared to the standard errors of the
(estimable) logistic regression coefficients. However, the MBLR and RLR
models as formulated will not protect against problems of estimability
in case \textit{every} response is quite sparse, because of the use of
an improper prior for the prior means $(A_{1}, \ldots, A_{G},
B_{0})$. If certain covariate or treatment categories are perfectly
correlated with every response, then one must either drop such
predictors or add additional response variables.

The Bayes factor for $\phi_{0} = (5, 5, 0.001, 0.001)$ can be computed
and compared to the 33 values found in the final grid of the Bayesian
estimation described above, which provides further evidence regarding
the prior standard deviations. In particular, large Bayes factors
against $\phi_{0}$ imply that the MBLR model fits the data better than
the RLR model, meaning that there is significant evidence that either
the responses have similar covariate profiles or that there are
significant covariate-by-treatment interactions.

\subsection*{Confidence Intervals for Odds Ratios}

Let the final estimate of, for example, $\beta_{gk}$ be $b_{gk}$,
so that the odds ratio point estimate is $\mathrm{OR}_{gk} = \exp(b_{gk})$.
Using the normal approximation to the posterior distribution of the
coefficients and the estimates of $V$ in equation (\ref{eq10}), 90\% confidence
intervals (posterior credible intervals) for the corresponding odds
ratios are given by
\begin{eqnarray}\label{eq18}
\hspace*{17pt}\mathrm{OR}.05 &=& \exp\bigl[b_{gk} - 1.645 \sqrt{v(\beta_{gk})}\bigr] < \mathrm{OR}\nonumber
\\[-8pt]
\\[-8pt]
& <&\exp\bigl[b_{gk} + 1.645 \sqrt{v(\beta_{gk})}\bigr] = \mathrm{OR}.95.\nonumber
\end{eqnarray}

For the main effects of covariates or for treatment, these provide
confidence (credible) intervals for odds ratios of the predictor vs the
response outcome. The odds ratio comparing two categories of a
multicategory covariate would be found by taking the ratio of the
corresponding exponentiated coefficients.

Interpreting the interaction effects of covariates with treatment arm is
tricky, since it would involve ratios of odds ratios. To aid in
interpretation, one can present in addition to the interaction
coefficients themselves, the sums of the treatment coefficient plus the
interaction coefficients. Confidence intervals for these sums are formed
in the usual way, taking into account the covariances between the
treatment coefficient and the interaction coefficients. When\break these sums
and their confidence limits are exponentiated, we get estimates and
limits for subgroup treatment-by-outcome odds ratios. These estimates
are oriented toward finding potentially vulnerable subgroups where the
adverse effect risk of treatment is especially high.

\section{Discussion of Methods and Alternate Models}

The philosophy of estimation is not to try to model the medical
mechanisms perfectly, but to provide a~reliable compromise between
pooling related sparse events in order to increase the sample size, and
fitting separate models to each event, with the corresponding loss of
power due to small samples. The selection of which issues to include in
an MBLR is important. There needs to be at least a superficial
plausibility that all or many of the selected outcome issues might have
similar odds ratios with treatment and with the covariates in the model,
what Bayesians call \textit{exchangeability}. Sometimes it may be
difficult to decide what other issues to include if attention has
focused primarily on a single and seemingly unique issue such as subject
death. Because it takes several degrees of freedom to estimate a
variance component, the values of some of the standard deviations in
equations (\ref{eq3})--(\ref{eq6}) may be poorly estimated if $K$ and/or $G$
 are not large, but
the use of Bayes factors and the computation of the $\pi_{s}$ in
(\ref{eq11}) and (\ref{eq16}) allow
some assessment and adjustment for this uncertainty.

The current model is quite similar in spirit to, and somewhat inspired
by, that proposed by Berry and Berry (\citeyear{r1}). They also assume that drug
adverse reactions are classified into similar medical groupings in
order
to use a shrinkage model to allow borrowing strength across similar
medical events.\vadjust{\goodbreak} They focus on treatment/comparator odds ratios only and
do not consider covariates or the use of logistic regression. They also
define a more complex model having many more variance components than
the one proposed here.

One might ask the question of why estimate covariate effects at all,
since in a randomized study the covariates should all be nearly
orthogonal to the treatment variable? The rationale in MBLR is not so
much to adjust for potential biases in the treatment main effect, but to
be able to include treatment-by-covariate interactions in order to
detect possibly vulnerable subgroups that might react differently to the
treatment. When $G$ is large (many covariate categories) it will often be
difficult to estimate so many parameters unless all the issues being
modeled occur frequently. The multiple comparisons involved make any
search for vulnerable subgroups difficult and subject to false alarms,
especially for sparse events. This makes the use of Bayesian shrinkage
of the interaction terms in (\ref{eq6}) especially valuable: it negotiates the
bias-variance trade-off among multiple event rates having possibly very
different sampling variances. Without this smoothing effect, estimates
of interactions affecting rare events will be so variable as to be
useless, which is why the RLR method is defined to estimate only main
effects.

The importance of avoiding undue rejection of the null hypothesis in the
presence of multiple post-hoc comparisons is central to being properly
conservative when evaluating treatment efficacy. There is a~question as
to how much this conservatism should extend to exploratory analyses of
safety issues. For example, the prior specification (\ref{eq6}) shrinks the
interaction prior means $B_{g}$ toward 0, whereas the main effect
prior means $A_{g}$ and $B_{0}$ are not shrunk toward 0. We prefer to
maintain maximum sensitivity to safety main effects, while accepting
that true interaction effects are less likely and need more false alarm
protection. We also encourage parallel computation of the
minimal-shrinkage regularized LR estimates discussed above, so that the
analyst can perform an easy comparison and sensitivity analysis of the
effects of shrinkage.

\begin{figure*}
\tablewidth=\textwidth
\tabcolsep=0pt
\begin{tabular*}{\textwidth}{@{\extracolsep{\fill}}ld{3.0}d{3.0}c@{}}
\hline
\textbf{Issue} & \multicolumn{1}{c}{\textbf{Treatment events}} & \multicolumn{1}{c}{\textbf{Comparator events}} & \multicolumn{1}{c}{\textbf{95\% C.I. for Odds Ratio}}
\\
\hline
Anuria & 8 & 0 & (1.0 , 295.4)
\\
Dry mouth & 308 & 65 & (3.9 , 6.7)
   \\
Hyperkalaemia & 218 & 162 & (1.1 , 1.7)
   \\
Micturition urgency & 13 & 3 & (1.2 , 12.6)
    \\
Nocturia & 19 & 7 & (1.1 , 6.1)
   \\
Pollakiuria & 193 & 34 & (4.1 , 8.5)
     \\
Polydipsia & 49 & 4 & (4.2 , 29.3)
    \\
Polyuria & 100 & 17 & (3.5 , 9.8)
    \\
Thirst & 543 & 66 & (7.5 , 12.6)
    \\
Urine output increased & 13 & 1 & (1.7 , 48.8)
 \end{tabular*}\vspace*{6pt}
 \begin{tabular*}{\textwidth}{@{\extracolsep{\fill}}lcc@{}}
Subject counts: & Treatment $=$ 3110 & Comparator $=$ 2642
\\
\hline
 \end{tabular*}
  \caption{Statistics for ten issues related to dehydration/renal
function for the pooled studies.}\label{dis1}
\vspace*{3pt}
\end{figure*}

The prior distributions in equations (\ref{eq3})--(\ref{eq6}) are all assumed to be normal
distributions. Many Bayesian researchers have pointed out that since
normal distributions generate few outliers, outliers may be
correspondingly suppressed under this assumption.\break Commonly suggested
alternative prior distributions are the double exponential and Student's
$t$, which tend to shrink outliers less. The double\vadjust{\goodbreak} exponential\break
(``lasso'') prior has nonstandard theoretical properties that make
computation of standard errors of coefficients problematical, and so
have been ruled out for this application. Alternative distributions like
Student's~$t$ are difficult to handle computationally in our complex
situation where there are hundreds of coefficients and multiple variance
components. The normal model that we use has a concave log posterior
density function and the iterative estimation algorithm is guaranteed to
converge.

There is a similar computational feasibility rationale for using the
discrete approximation to the\break distribution of prior standard deviations.
It is more common in the recent Bayesian literature to use Gibbs
sampling or another Markov chain Monte\break Carlo (MCMC) method to estimate
the posterior distributions of all parameters. Two reasons for
preferring to avoid such methods are as follows: first, we want to allow
scientists without much statistical sophistication, much less experience
with fancy Bayesian computational methods, to use MBLR and these users
would have trouble assessing convergence of such high-dimensional MCMC
runs. Second, these users might also be uncomfortable with the fact that
repeating an analysis on the same data typically leads to slightly, but
noticeably, different answers. The method for handling the variance
component estimation outlined above provides computationally and
statistically reliable answers within a feasible computational burden.
As described above, there are three roughly equally expensive stages in
the model fitting computations: the two preparatory\break stages of finding
the maximum of the posterior distribution and then evaluating it on an
initial grid to find scale parameters in each direction, and the last
stage of evaluating the model on the final grid to approximate the
posterior distribution of the variance components.

\section{Example Analysis}\label{sec5}

\subsection*{Data Description}

The data used for the example analyses are from a pool of eight studies,
kindly contributed by an anonymous partner. Four of the studies were for
one indication and four were for a second indication. There were a total
of 5752 subjects in the pooled studies, 3110 in the Treatment arm and
2642 in the Comparator arm.

Display~\ref{dis1} shows statistics from these studies for a~set of ten issues
related to dehydration and/or renal function. All ten issues show up
with greater frequency in the treatment arm than in the comparator. The
final two columns are the endpoints of 95\% confidence intervals for the
odds ratios comparing treatment and comparator groups in the pooled
data, computed using a normal approximation for the log (odds ratio)
after adding 0.5 to every cell of each $2\times 2$ table. It is clear that many
of these issues are associated with treatment, and we wish to
investigate the commonality of these medically related issues, as well
as the possibility that certain subgroups of subjects may be more or
less affected by these associations.

\begin{figure}
\tablewidth=240pt
\tabcolsep=0pt
\begin{tabular*}{240pt}{@{\extracolsep{\fill}}ld{4.0}d{4.0}}
\hline
 & \multicolumn{1}{c}{\textbf{Treatment}} & \multicolumn{1}{c}{\textbf{Comparator}}
 \\
 \hline
Gender ${=}$ F & 908 & 685
\\
Gender ${=}$ M & 2202 & 1957
\\
Study ${=}$ A1 & 246 & 84
\\
Study ${=}$ A2 & 120 & 120
\\
Study ${=}$ A3 & 239 & 80
\\
Study ${=}$ A4 & 191 & 63
\\
Study ${=}$ B1 & 102 & 103
\\
Study ${=}$ B2 & 17 & 11
\\
Study ${=}$ B3 & 123 & 120
\\
Study ${=}$ B4 & 2072 & 2061
\\
Renal history ${=}$ Y & 190 & 191
\\
Renal history ${=}$ N & 2920 & 2451
\\
Age ${=}$ 50 or under & 382 & 348
\\
Age ${=}$ 51 to 65 & 1089 & 902
\\
Age ${=}$ 66 to 75 & 948 & 820
\\
Age ${=}$ Over 75 & 691 & 572
\\[6pt]
All patients & 3110 & 2642
\\
\hline
\end{tabular*}
  \caption{Distribution of subjects by covariates and treatment arm.}\label{tab1}
\end{figure}

Display~\ref{tab1} shows the four covariates selected as grouping variables for
this analysis: Gender, Study ID, Renal History and Age. Recall that of
the 8 studies being pooled, there were 4 studies for each of two
potential indications for the drug. The Study ID values of A1--A4 and
B1--B4 distinguish the studies for indication A and indication B. The
Renal History variable distinguishes those subjects whose medical
history (before randomization) includes one or more renal problems. As
can be seen from Display~\ref{tab1}, there are many more male than female
subjects, and the age range 51 to 75 predominates. Three of the studies
for indication A had about a 3:1 split of Treatment to Comparator
subject counts, while the other five studies are more equally split.
Study B2 had only 28 subjects total, while Study B4 had 4133 subjects,
over two-thirds of the total in the pool. Only about 7\% of the subjects
had a previous history of renal problems.

Display~\ref{dis1} shows that five of the ten issues affected fewer than 10
Comparator-group subjects, whereas there are 16 separate covariate
groups in Display~\ref{tab1}. This makes it unlikely that those rare issues would
occur in every treatment--covariate combination, which is necessary for
convergence of a~standard LR where the model includes all
treatment--covariate interaction terms. In fact, only 3 of the 10 issues
satisfy this condition, confirming the necessity of some special
technique such as MBLR to try to estimate treatment-by-covariate
interactions, and, in fact, even a main-effects only model would not be
estimable by standard logistic regression applied to the rarer of these
response issues, making the regularized LR necessary for this
example.\looseness=1

\begin{figure*}
\tablewidth=\textwidth
\tabcolsep=0pt
\begin{tabular*}{\textwidth}{@{\extracolsep{\fill}}ld{1.3}d{1.3}d{1.3}d{1.3}c@{}}
\hline
 & \multicolumn{1}{c}{$\boldsymbol{\sigma_{A}}$} &
 \multicolumn{1}{c}{$\boldsymbol{\sigma_{0}}$} &
 \multicolumn{1}{c}{$\boldsymbol{\sigma_{B}}$} &
 \multicolumn{1}{c}{$\boldsymbol{\tau}$} &
 \multicolumn{1}{c}{$\boldsymbol{\mathit{PROB}}$}
 \\
 \hline
\phantom{0}0 & 5.000 & 5.000 & 0.001 & 0.001 & \phantom{0}0.00\%\\
\phantom{0}1 & 0.327 & 0.688 & 0.161 & 0.196 & 15.90\%\\
\phantom{0}2 & 0.276 & 0.505 & 0.110 & 0.312 & \phantom{0}1.87\%\\
\phantom{0}3 & 0.276 & 0.505 & 0.232 & 0.118 & \phantom{0}3.02\%\\
\phantom{0}4 & 0.276 & 0.879 & 0.110 & 0.118 & \phantom{0}2.71\%\\
\phantom{0}5 & 0.276 & 0.879 & 0.232 & 0.312 & \phantom{0}4.32\%\\
\phantom{0}6 & 0.384 & 0.505 & 0.110 & 0.118 & \phantom{0}4.25\%\\
\phantom{0}7 & 0.384 & 0.505 & 0.232 & 0.312 & \phantom{0}1.83\%\\
\phantom{0}8 & 0.384 & 0.879 & 0.110 & 0.312 & \phantom{0}5.01\%\\
\phantom{0}9 & 0.384 & 0.879 & 0.232 & 0.118 & \phantom{0}1.75\%\\
10 & 0.252 & 0.423 & 0.090 & 0.091 & \phantom{0}0.14\%\\
11 & 0.252 & 0.423 & 0.276 & 0.387 & \phantom{0}0.42\%\\
12 & 0.252 & 0.969 & 0.090 & 0.387 & \phantom{0}0.88\%\\
13 & 0.252 & 0.969 & 0.276 & 0.091 & \phantom{0}0.83\%\\
14 & 0.416 & 0.423 & 0.090 & 0.387 & \phantom{0}0.51\%\\
15 & 0.416 & 0.423 & 0.276 & 0.091 & \phantom{0}0.10\%\\
16 & 0.416 & 0.969 & 0.090 & 0.091 & \phantom{0}5.40\%\\
17 & 0.416 & 0.969 & 0.276 & 0.387 & \phantom{0}0.13\%\\
18 & 0.231 & 0.688 & 0.161 & 0.196 & \phantom{0}1.86\%\\
19 & 0.448 & 0.688 & 0.161 & 0.196 & \phantom{0}2.13\%\\
20 & 0.327 & 0.350 & 0.161 & 0.196 & \phantom{0}1.37\%\\
21 & 0.327 & 1.053 & 0.161 & 0.196 & \phantom{0}6.96\%\\
22 & 0.327 & 0.688 & 0.074 & 0.196 & \phantom{0}8.12\%\\
23 & 0.327 & 0.688 & 0.327 & 0.196 & \phantom{0}0.75\%\\
24 & 0.327 & 0.688 & 0.161 & 0.070 & \phantom{0}5.80\%\\
25 & 0.327 & 0.688 & 0.161 & 0.473 & \phantom{0}3.21\%\\
26 & 0.192 & 0.688 & 0.161 & 0.196 & \phantom{0}0.87\%\\
27 & 0.518 & 0.688 & 0.161 & 0.196 & \phantom{0}2.43\%\\
28 & 0.327 & 0.232 & 0.161 & 0.196 & \phantom{0}0.02\%\\
29 & 0.327 & 1.196 & 0.161 & 0.196 & \phantom{0}4.69\%\\
30 & 0.327 & 0.688 & 0.049 & 0.196 & \phantom{0}5.54\%\\
31 & 0.327 & 0.688 & 0.447 & 0.196 & \phantom{0}0.00\%\\
32 & 0.327 & 0.688 & 0.161 & 0.041 & \phantom{0}6.18\%\\
33 & 0.327 & 0.688 & 0.161 & 0.670 & \phantom{0}1.01\%  \\[6pt]
\multicolumn{1}{l}{Mean} & 0.336 & 0.756 & 0.146 & 0.196 & \phantom{0}6.39\%\\
\multicolumn{1}{l}{St.Dev.} & 0.053 & 0.183 & 0.053 & 0.105 & \\
\hline
\end{tabular*}
\caption{Calculation summary for the final grid of prior standard
deviations.}\label{tab2}
\end{figure*}

\subsection*{Posterior Distributions for Prior Standard Deviations}

This example has $K = 10$, $J = 4$ and $G = 16$, with the total number of
parameters (elements of $\theta$) to estimate being $M = 2(G+1)(K+1) - 1
= 373$, with $M^* = 285$ degrees of freedom. Display \ref{tab2} shows various results
as a function of the four prior standard deviations. The top row 0
describes the regularized LR case where $\sigma_{A} = 5$,
$\sigma_{0} = 5$, $\sigma_{B} = 0.001$, $\tau = 0.001$. The rows labeled
1--33 in Display~\ref{tab2} show results for the final grid used to approximate
the posterior distribution of $\phi$. The row 1 values are the maximum
posterior estimates (transformed from the scale of $\lambda$ to that of
$\phi$) estimated by the final response surface fit described above.
Rows 2--33 show the remaining values of the final stage grid. In this
example, all stages of the estimation required a total of about 400
iterations through the data, that is, about 400 evaluations of (\ref{eq8}) and
its first and second derivatives with respect to $M^* = 285$
parameters.\looseness=1

The rightmost column in Display~\ref{tab2}, headed ``PROB,'' shows the values of
$100\pi\%$, as defined by (\ref{eq16}). As discussed above, these
probabilities have been adjusted so that the discrete distribution of
$\lambda$ matches the means and variances of the continuous distribution
of $\lambda$ as estimated by the response surface fit to the values of
$\log g$.

The bottom two rows of Display~\ref{tab2} show the posterior mean and standard
deviations of the components of $\phi$ using this 33-point discrete
approximation. It can be seen that the values are approximately
($\sigma_{A}=0.34$, $\sigma_{0}=0.76$, $\sigma_{B}=0.15$, $\tau
=0.20$). The value in the row marked ``Mean'' and the column marked
``PROB'' is computed as $\sum_{s}\pi_{s}^{2} = 0.0639$,
which is a measure of the dispersion of the probabilities $\pi_{s}$.
The smaller it is, the more spread out are the probabilities among the
33 grid points. Large values of $\sum_{s}\pi_{s}^{2}$,
say, values above 0.2, would imply that the scale or location of the
grid might be poorly chosen, so that only a few points on the grid are
very probable.

\begin{figure*}

\includegraphics{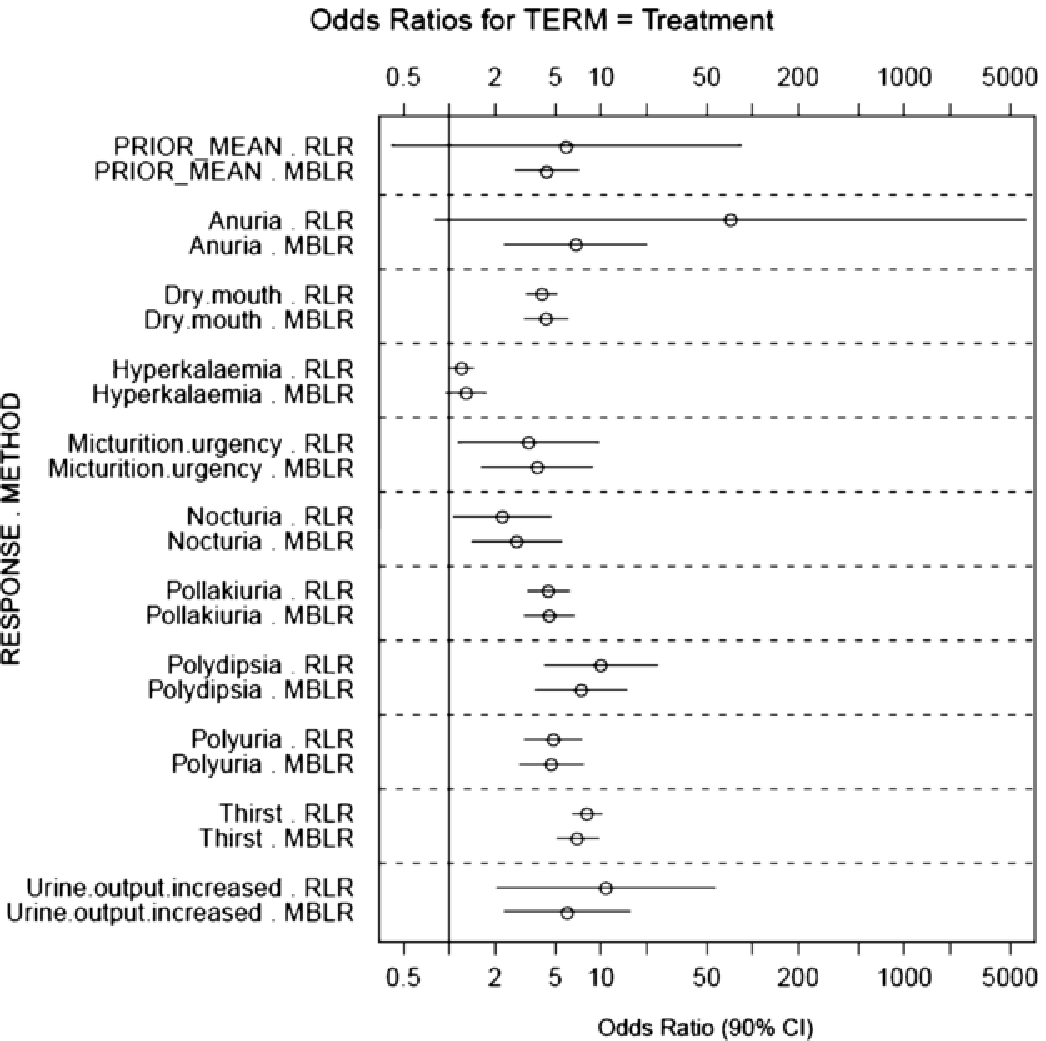}

  \caption{Estimates of main effect of treatment by method and response
variable.}\label{dis4}
\end{figure*}

\subsection*{Comparison of MBLR and RLR Estimates of Treatment Effects}

Display~\ref{dis4} shows estimation results for the treatment main effects for
each of the two methods and for each response event and for the prior
mean of all responses. The prior mean odds ratio is defined as
$\exp(B_{0})$, whereas the treatment odds ratio for the $k$th
response is $\exp(\beta_{0k})$. For each\vadjust{\goodbreak} combination the odds ratio
and its approximate 90\% confidence (credible) interval are shown, based
on (\ref{eq18}). Comparing the MBLR to the RLR estimates, we see that the MBLR
estimates are pulled away from the RLR estimates and ``shrunk'' toward
the MBLR prior\break mean, which represents the average or typical odds ratio
across response issues. The degree of shrinkage is greatest for the
highest-variance RLR estimates, corresponding to the rare issues such as
Anuria and Urine output increased. For these two issues, although the
MBLR odds ratio estimate is smaller than the corresponding RLR odds
ratio, but so are their posterior variances, so that the lower bounds of
the MBLR intervals are greater, providing greater statistical
significance from the null hypothesis of $\mathrm{OR} = 1$. Even though all 8
occurrences of Anuria were in the treatment arm, the treatment effect
does not show up as significant with the multiple-predictor RLR
model---the MBLR estimate of the effect on Anuria seems more reasonable.

Inspection of Display~\ref{dis4} shows that not all of the MBLR confidence
intervals are narrower than the corresponding RLR interval. The reverse
is true for the more frequent responses such as Hyperkalaemia and
Thirst. In these cases, the MBLR estimates do not benefit much from the
relatively weak prior distribution, and their posterior variances are
adversely impacted by the uncertainty in the variance component
estimation as well as the need to estimate all of the interaction
parameters, which are assumed away by the RLR model.

\begin{figure*}

\includegraphics{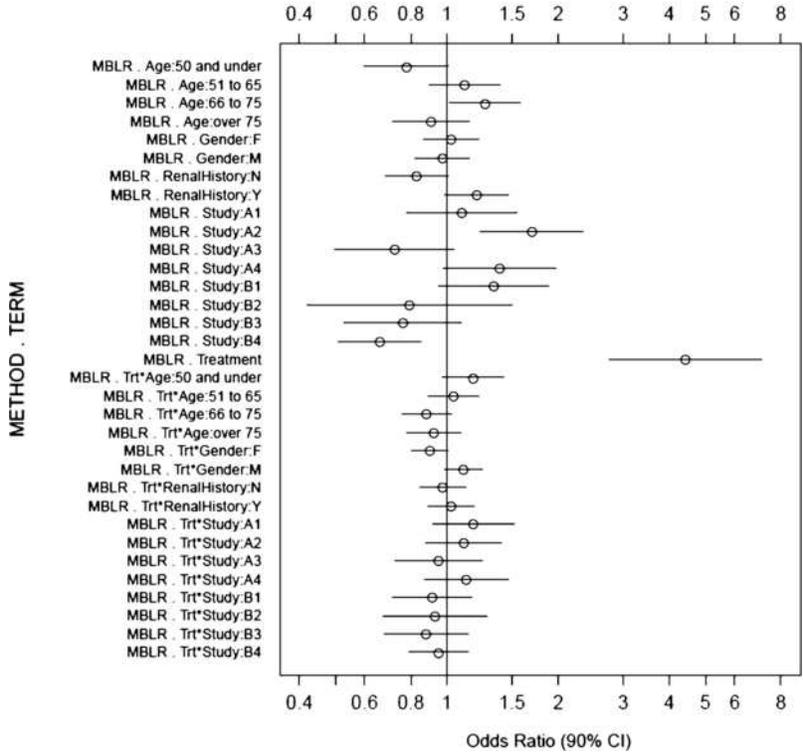}

  \caption{Estimates of PRIOR\_MEAN from MBLR.}\label{dis5}
  \end{figure*}

\subsection*{MBLR Estimates of Prior Means}

Display~\ref{dis5} graphs the MBLR estimates of the (exponentiated) prior means
$\{A_{g}, B_{0}, B_{g}\}$, with their 90\% CIs. These are
interpreted as effects for a ``typical'' response variable. Remembering
that coefficients for categories of each covariate must sum to~0, the
corresponding odds ratios must average to~1 when plotted on a log scale.
The middle interval shows the main effect of treatment, the intervals
above show covariate main effects, and the intervals below show
treatment interactions. As also shown in Display~\ref{dis4}, the treatment effect
prior mean is about 4.4 on the odds ratio\vadjust{\goodbreak} scale, with 90\% limits of
(2.7, 7.1). The main effects of covariate estimates, shown above the
treatment line, can be thought of as the effects of covariate categories
within the comparator arm, and as centers of shrinkage across the
responses. Thus, the rates of these events in the comparator arm are
somewhat less for Age:50 and under and for Renal History:N. Also,
Study:A2 had a~particularly high event rate, while Study:B4 had
a~particularly low event rate. But none of these differences in groups
based on covariates are as large as the treatment effect.

The lower set of estimates in Display~\ref{dis5} portray the treatment--covariate
interactions. As can be seen, these effects are smaller than the main
covariate effects and much smaller than the main treatment effect. The
treatment effect estimates within the four studies for Indication A are
all larger than the four estimates for the Indication B studies, but the
uncertainty intervals all overlap considerably. Although this does not
rule out larger interaction effects for some of the response variables,
the fact that $\sigma_{A}$, is about 0.3 and both $\sigma_{B}$,
and $\tau$ are each less than 0.2 means that such effects for individual
responses are also likely to be fairly small. Since $\sigma_{0}$ is
about 0.76,\vadjust{\goodbreak} there is more room for variation in treatment main effect
among the responses, as we also saw in Display~\ref{dis4}, where the Treatment
odds ratios ranged from 1.3 for Hyperkalaemia to 7.4 for
Polydipsia.

\begin{figure*}

\includegraphics{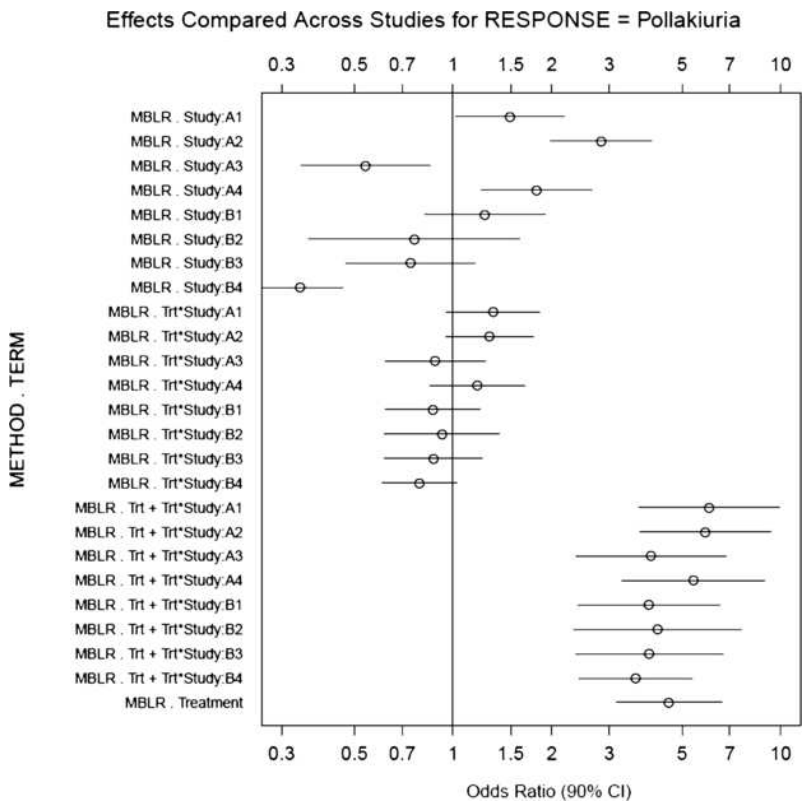}

  \caption{MBLR estimates of odds ratios relating to the Study covariate
for the response Pollakiuria.}\label{dis6}
 \end{figure*}

The prior means of the treatment by covariate interactions (the bottom
16 intervals of Display~\ref{dis5}) have especially small posterior means, as
might be expected given that they have been shrunk toward 0 because of
the small value of $\tau$, with posterior mean $=$ 0.196 in Display
\ref{tab2}.
Another way of saying this is that the estimates of $B_{g}$ were so
small compared to their sampling variances that only a small value of
$\tau$ is compatible with these results and the assumption of (\ref{eq6}).

The estimates of prior means under the regularized LR model are less
interesting. Assuming that~$\sigma_{A}$ and $\sigma_{0}$ are large
implies that $A_{g}$ and $B_{0}$ cannot be estimated well and will
thus have wide confidence intervals, and of course assuming no
interactions means that the $B_{g} = 0$ for $g > 0$.

\subsection*{Breakdown of Estimates by Study for Issue Pollakiuria}

Display~\ref{dis6} shows the MBLR 90\% intervals for odds ratios relating to the
Study ID covariate and the issue Pollakiuria (very frequent daytime
urination). The $2\times 2$ table information in Display~\ref{dis1} shows that this was
highly associated with treatment (193:34 split by treatment:comparator).
Our discussion focuses on whether and how the results differ by the
studies being pooled, and what summary conclusions are justified across
studies. The goal is similar to meta-analysis, except that we have
complete data from each study and so can adjust for more potentially
biasing between-study differences.

The top eight intervals in Display~\ref{dis6} show the Study main effects,
corresponding to relative differences among the comparator arm odds of
reporting Pollakiura within the studies. These differential estimates
are adjusted for the other covariates Age, Gender and RenalHistory.
There are relatively large and significant study effects, especially
between\break Study~A2 and Study B4, where the estimated odds ratio is over 8
(2.8 versus 0.34 on the horizontal axis), with relatively narrow 90\%
intervals.

The next set of eight intervals shows the Treatment by Study interaction
estimates. Although the differences are not as large as in the
comparator arms, the pattern is similar, in that the studies that had a
large base rate of Pollakiuria tended to have larger increases in\vadjust{\goodbreak}
adjusted Pollakiura rates. The three studies having the largest
treatment effects (A1, A2, A4) are all based on Indication A. These
estimates are somewhat hard to interpret, being ratios of odds ratio
estimates. The lower set of intervals return to the simple odds ratio
scale by adding (on the log scale) the interaction estimates to the main
effect of treatment. The very bottom interval shows the 90\% interval
for the treatment main effect, and the central points for the eight
intervals above it average to the center point at the
bottom.

These last 9 intervals in Display~\ref{dis6} are reminiscent of the way a
meta-analysis is often presented~in a~``ladder plot,'' with estimates of
effect for each~study, and followed by a combined treatment estimate at
the bottom. However, there are certain differences due to the more
complex MBLR model. First, as mentioned above, these estimates have been
adjusted for differential covariate distributions across studies.
Second, the Pollakiuria estimates here have been shrunk toward the prior
mean estimates of the odds ratios involving all responses. Third, the
shrinkage of interaction estimates toward 0, governed by $\tau$ in (\ref{eq6}),
is similar to the shrinkage toward a common mean effect that occurs in a
random effects meta-analysis. Fourth, the weight that each study\vadjust{\goodbreak}
contributes to the overall estimate is governed by a more complex
formula than in either the standard fixed or random effects
meta-analyses. However, it does share with the random effects
methodology the fact that relative weights are much attenuated compared
to relative sample sizes. Finally, this more complex calculation means
that the single-study treatment estimates in the above MBLR graph do not
preserve the between-study differences, as might be shown in a standard
meta-analysis presentation.

The response Pollakiuria was chosen as the example for Display~\ref{dis6} because
that issue showed a greater Treatment-by-Study effect than other issues:
for example, in Display~\ref{dis6} the Trt\tsup{*}Study:A1 effect is 1.33, while the
Trt\tsup{*}Study:B4 effect is 0.79, for a ratio of 1.68, and the two 90\%
intervals barely overlap. Is this post-hoc selection legitimate?
Clearly, this way of finding ``interesting'' results is biased in many
standard settings. However, the Bayesian shrinkage methodology tends to
offset such biases, as will be seen in the simulation results to follow.

\section{Simulation Study of MBLR and RLR}\label{sec7}

The statistical properties of MBLR are studied using a simulation of the
model that MBLR assumes. The purpose is to compare the accuracy of the
MBLR results with that of the RLR results in the context of a situation
like that in the example of Section~\ref{sec5}, where there are rare
events and sparse data. The simulation emulates that example in the
sense that the distribution of subject covariates and treatment
assignment matches the data in Section~\ref{sec5} exactly. Also, the list of
response issues is the same and the baseline probabilities (as measured
by the intercept term in the logistic regressions) of each response in
the simulation are similar to that in the data of Section~\ref{sec5}. The
protocol for each simulation involves the following steps:
\begin{enumerate}[13.]
\item Set the $K$ intercept term values $\alpha_{0k}$, one for each of
the responses.

\item Set the $G+1$ prior means $A_{1}, A_{2}, \ldots, A_{G}, B_{0}$.

\item Set the four prior standard deviations $\phi = (\sigma_{A},\break
\sigma_{0}, \sigma_{B}, \tau)$.

\item Repeat steps (5 through 12) $N_{\mathrm{SIM}}$ times:
\begin{enumerate}[12.]
\item[5.] Draw $\{\alpha_{gk}\}$ from $N(A_{g}, \sigma_{A}^{2})$,
$g = 1, \ldots, G$; $k = 1,\ldots, K$.

(Note: all random variable generation is performed using built-in $R$
functions. Also, constraints that $\alpha_{gk}$ must sum to 0 as $g$
varies over the categories of each single covariate are enforced by
subtracting means over the corresponding covariate from the originally
drawn $\alpha_{gk}$. An analogous procedure is used in steps 7 and
8.)

\item[6.] Draw $\{\beta_{0k}\}$ from $N(B_{0}, \sigma_{0}^{2})$, $k =
1,\ldots,
K$.

\item[7.] Draw $\{B_{g}\}$ from $N(0, \tau^{2})$, $g = 1,\ldots, G$.

\item[8.] Draw $\{\beta_{gk}\}$ from $N(B_{g}, \sigma_{B}^{2})$,
$g = 1, \ldots, G$; $k = 1, \ldots, K$.

\item[9.] For each set of $n_{i}$ subjects having the same covariate values
and treatment assignment, compute $Z_{ik}$ and $P_{ik}$ using (\ref{eq1})
and (\ref{eq2}), $i = 1, \ldots, m$; $k = 1, \ldots, K$.

\item[10.] Draw $\{N_{ik}\}$ from binomial $(n_{i}, P_{ik})$, $i =\break 1,
\ldots, m$; $k = 1, \ldots, K$.

\item[11.] Fit both the MBLR and the RLR model to the counts $\{N_{ik}\}$.

\item[12.] Update cumulative summaries of estimation results for each
simulation as described below.
\end{enumerate}
\item[13.] Create reports summarizing the estimation accuracy of the two
methods regarding all parameters.
\end{enumerate}

\subsection*{Simulation Summary Statistics}

There are $M = 2(G + 1)(K + 1) - 1$ parameters being estimated and two
estimation methods: MBLR and RLR,\vadjust{\goodbreak} so the total number of estimators
being evaluated is $R = 2M$. For simulation $s$ ($s = 1, \ldots,\break
N_{\mathrm{SIM}}$) and for estimate $r$ ($r = 1, \ldots, R$), let:
\begin{enumerate}
\item[] $\theta_{rs} = \mathrm{true}$ value of parameter $r$ for simulation $s$, as
defined by steps 1, 2, 5, 6, 7 and 8,

\item[] $q_{rs} = \mathrm{estimated}$ value (posterior mean) of parameter $r$ for
simulation $s$,

\item[] $\mathit{se}_{rs} = \mathrm{estimated}$ SE (posterior standard deviation) for parameter
estimate $r$ for simulation $s$,

\item[] $\mathrm{BIAS}_{r} = \sum_{s}(q_{rs} - \theta_{rs}) /
N_{\mathrm{SIM}}$ [average estimation error],

\item[] $\mathrm{RMSE}_{r} = \sqrt{(\sum_{s}(q_{rs} -
\theta_{rs})^{2} / N_{\mathrm{SIM}})}$ [square root of mean squared
estimation error],

\item[] $Z^{2}_{r} = (\sum_{s}(q_{rs} -
\theta_{rs})^{2}/\mathit{se}_{rs}^{2}) / N_{\mathrm{SIM}}$ [average
squared standardized estimation error],

\item[] $\mathrm{CI}.05_{r} = (\# \mbox{ times } q_{rs} + 1.645 \mathit{se}_{rs} <
\theta_{rs}) / N_{\mathrm{SIM}}$\break [\mbox{proportion} of times 90\% CI is too
low],

\item[] $\mathrm{CI}.95_{r} = (\# \mbox{ times } q_{rs} - 1.645 \mathit{se}_{rs} >
\theta_{rs}) / N_{\mathrm{SIM}}$\break [\mbox{proportion} of times 90\% CI is too
high].
\end{enumerate}

These summary statistics focus on the estimation accuracy of $q_{rs}$
and also on the calibration accuracy of $\mathit{se}_{rs}$. We want BIAS and
RMSE to be as close to 0 as possible, we want $Z^{2}$ to be near 1, and
we want CI.05 and CI.95 to be near 0.05. The~$R$ estimates can be grouped
by the two methods, the ($K + 1$) responses (counting PRIOR\_MEAN as
a~generalized response) and the $2G + 2$ different term definitions. The
term definitions fall into three general \textit{term types}:
\begin{eqnarray*}
\mathrm{COV} &=& \{A_{g}, \alpha_{gk}\},
\\
\mathrm{TREAT} &=& \{B_{0}, \beta_{0k} \},
\\
\mathrm{TRT^{*}COV} &=& \{B_{g}, \beta_{gk}\}.
\end{eqnarray*}

We can summarize the simulation of the $R$ estimates by averaging the six
accuracy summaries listed above over groups defined by method, response
and/or term type.

Finally, for the MBLR, we can summarize the posterior means and standard
deviations of the four estimated prior standard deviations.

\subsection*{Simulation Design}

The simulations are designed to compare variations in three design
factors, each at two levels, so that 8 separate simulations were
performed, with each simulation having $N_{\mathrm{SIM}} = 250$ replications,
and so that a simple comparison of the two levels of each factor will be
based on 1000 replications at each level. The three design factors
correspond to two different choices at each of the first three steps in
the simulation protocol given above:
\begin{enumerate}[\quad]
\item[Factor 1, Level 1:] Frequent responses only (most frequent 5 in the
example data).
\item[Factor 1, Level 2:] Both frequent and rare responses (all 10 issues in
the example data).

\begin{figure}
\tablewidth=240pt
\tabcolsep=0pt
\begin{tabular*}{240pt}{@{\extracolsep{\fill}}lcc@{}}
\hline
\multicolumn{1}{l}{\textbf{Response}} & \multicolumn{1}{l}{\textbf{Intercept}} & \multicolumn{1}{l}{\textbf{Base.Prob}}
\\
\hline
\phantom{0}1. Hyperkalaemia & $-$2.906 & 0.0547\\
\phantom{0}2. Thirst & $-$3.333 & 0.0357\\
\phantom{0}3. Dry mouth & $-$3.429 & 0.0324\\
\phantom{0}4. Pollakiuria & $-$3.645 & 0.0261\\
\phantom{0}5. Polyuria & $-$4.787 & 0.0083\\[3pt]
\phantom{0}6. Nocturia & $-$5.646 & 0.0035\\
\phantom{0}7. Polydipsia & $-$5.819 & 0.0030\\
\phantom{0}8. Micturition urgency & $-$6.618 & 0.0013\\
\phantom{0}9. Urine output increased & $-$6.972 & 0.0009\\
10. Anuria & $-$7.722 & 0.0004\\
\hline
\end{tabular*}
\caption{Estimated values of intercept terms $\alpha_{0k}$ that
are used in the simulations. When $K = 5$, only the first 5 responses in
the display are used. The baseline probabilities are defined as
$\exp(\alpha_{0k})$, which range from 0.055 for Hyperkalaemia to
0.00044 for Anuria.}\label{tab3}
\end{figure}

The $K = 10$ situation uses the same 10 issues as in the example of
Section~\ref{sec5}, with the values of the intercept terms $\alpha_{0k}$ set
equal to the estimated values from the real data, shown in Display~\ref{tab3}.
The baseline probabilities are defined as $\exp(\alpha_{0k})$, also
shown, which range from 0.055 for Hyperkalaemia to 0.00044 for Anuria.
When $K = 5$, the most frequent 5 response issues are used, as shown in
rows 1--5 of Display~\ref{tab3}.

\item[Factor 2, Level 1:] Average of main $\mathrm{effects} = 0$ ($A_{g} = 0$ for all $g$,
$B_{0} = 0$).
\item[Factor 2, Level 2:] Prior Means of main effects relatively large nonzero
values.
\begin{figure}
\tablewidth=240pt
\tabcolsep=0pt
\begin{tabular*}{240pt}{@{\extracolsep{\fill}}ld{2.3}cd{2.3}@{}}
\hline
\textbf{Term} & \multicolumn{1}{c}{\textbf{Estimated}} & \textbf{Level 1} & \multicolumn{1}{c}{\textbf{Level 2}}\\
\hline
Gender: F & 0.028 & 0 & 0.056\\
Gender: M & -0.028 & 0 & -0.056\\
Study: A1 & 0.094 & 0 & 0.188\\
Study: A2 & 0.529 & 0 & 1.058\\
Study: A3 & -0.325 & 0 & -0.65\\
Study: A4 & 0.33 & 0 & 0.66\\
Study: B1 & 0.293 & 0 & 0.586\\
Study: B2 & -0.232 & 0 & -0.464\\
Study: B3 & -0.273 & 0 & -0.546\\
Study: B4 & -0.417 & 0 & -0.834\\
RenalHistory: N & -0.187 & 0 & -0.374\\
RenalHistory: Y & 0.187 & 0 & 0.374\\
Age: 50 and under & -0.251 & 0 & -0.502\\
Age: 51--65 & 0.109 & 0 & 0.218\\
Age: 66--75 & 0.239 & 0 & 0.478\\
Age: over 75 & -0.097 & 0 & -0.194\\
Treatment & 1.484 & 0 & 2.968\\
\hline
\end{tabular*}
\caption{Prior means for the main effects, as estimated by MBLR from
the real data, and as varied in the simulations, either all zeros (Level
1), or set to twice the estimated values (Level 2).}\label{tab4}
\end{figure}
\begin{figure*}
\tablewidth=\textwidth
\tabcolsep=0pt
\begin{tabular*}{\textwidth}{@{\extracolsep{\fill}}lcccccccc@{}}
\hline
 & $\boldsymbol{\sigma_{A}}$  &
 $\boldsymbol{\mathrm{SD}\sigma_{A}}$ &
 $\boldsymbol{\sigma_{0}}$  &
 $\boldsymbol{\mathrm{SD}\sigma_{0}}$ &
 $\boldsymbol{\sigma_{B}}$  &
 $\boldsymbol{\mathrm{SD}\sigma_{B}}$ &
 $\boldsymbol{\tau}$ & $\boldsymbol{\mathrm{SD}_{\tau}}$
 \\
 \hline
All MBLR simulations & 1.035 & 0.130 & 1.005 & 0.271 & 0.988 & 0.236 & 1.088 & 0.368\\
Responses: Frequent & 1.044 & 0.144 & 1.004 & 0.283 & 1.004 & 0.253 & 1.073 & 0.373\\
Responses: Freq ${+}$ Rare & 1.026 & 0.116 & 1.007 & 0.260 & 0.971 & 0.219 & 1.102 & 0.363\\
Mean effects: Zero & 1.034 & 0.133 & 1.005 & 0.276 & 1.000 & 0.245 & 1.091 & 0.376\\
Mean effects: Large & 1.035 & 0.126 & 1.006 & 0.267 & 0.975 & 0.227 & 1.084 & 0.360\\
Prior SDs: Small & 1.037 & 0.153 & 1.131 & 0.353 & 0.954 & 0.340 & 1.136 & 0.476\\
Prior SDs: Large & 1.033 & 0.106 & 0.879 & 0.190 & 1.022 & 0.132 & 1.039 & 0.259\\
\hline
\end{tabular*}
\caption{Summary of estimation of prior standard deviations (PSD) in
the MBLR simulations. All estimated PSDs are divided by the true PSD to
put their sampling distributions on a common scale. The row ``All
Simulations'' shows means and standard deviations of normalized
estimates across all 2000 simulations. Other rows show results for
subsets of 1000 simulations broken down by the two levels of each of
the three design factors in the experiment. See text for explanation of
the design factors and their levels.}\label{tab5}
\end{figure*}

For Level 2, the values of $A_{1}, \ldots, A_{G}, B_{0}$ are set to
two times the values estimated in the analysis of the actual data.
Display~\ref{tab4} shows the coefficient values as estimated and as used in the
simulations.

\item[Factor 3, Level 1:] $\sigma_{A} = 0.4$, $\sigma_{0} = 0.6$,
$\sigma_{B} = 0.2$, $\tau = 0.2$ (Small PSDs).
\item[Factor 3, Level 2:] $\sigma_{A} = 1.0$, $\sigma_{0} = 1.2$,
$\sigma_{B} = 0.8$, $\tau = 0.8$ (Large PSDs).

The Level 1 values of prior standard deviations are similar to those
estimated from the example data, while the Level 2 values are
significantly larger.
\end{enumerate}

All simulations create 5752 subjects having the same joint distribution
of covariates and treatment allocations as the actual data and as
summarized in Display~\ref{tab1}. Thus, $G = 16$ and $M = 203$, $R = 566$ when $K = 5$,
while $M = 373$, $R = 1066$ when \mbox{$K = 10$}.

\subsection*{Simulation Results}

Display~\ref{tab5} shows summaries of the distributions of (square roots of)
variance component estimates, which are denoted PSDs for prior standard
deviations in the model equations (\ref{eq3})--(\ref{eq6}). Since there are 4 separate PSDs
in the model, and the simulations are run at two sets of PSDs, the
scales of all the PSDs in Display~\ref{tab5} have been normalized by dividing
each estimated PSD and each estimated sampling standard deviation by the
true PSD used in the corresponding simulation. Thus, a~value of 1 for an
average estimated PSD in Display~\ref{tab5} is interpreted as an unbiased
estimate, and a value of 0.1 for the standard deviation of the sampling
distribution of a PSD in Display~\ref{tab5} is interpreted as a coefficient of
variation of 10\%.

Display~\ref{tab5} has 8 columns and 7 rows. There are 4 pairs of columns,
corresponding to the sampling means and standard deviations of the
estimates of each of the four PSDs in the model. The 7 rows of Display~\ref{tab5}
correspond to different subsets of the 2000 simulations. Row 1 shows
averages over all simulations, whereas the other rows show averages over
a~subset of 1000 simulations corresponding to the levels of each of the
factors in the experimental design. For example, consider the columns
labeled $\sigma_{0}$ and $\mathrm{SD}\sigma_{0}$ in Display~\ref{tab5}. In row 1, 1.005
implies that overall the mean of estimates of the Treatment PSD are
within 0.5\% of the true value, and the next value of $\mathrm{SD}\sigma_{0} =
0.271$ implies that individual estimates are typically about 27\% off
the true value. Of course that value is principally reflective of the
sample size and the experimental design of the clinical studies. All
simulations used the same clinical study setups, but there was variation
according to the three factors in the simulation. Going down the rows in
these same two columns, we see that estimates of $\sigma_{0}$ had almost
exactly the same means and standard deviations whether all ten responses
were being simulated or whether just the most frequent five responses
were simulated. Similarly, the next two rows show that there was
virtually no difference in the sampling means and standard deviations of
$\sigma_{0}$ between the situation where the average effects are about 0
versus relatively large effects. However, the final two rows of the
Display show that when all four PSDs are small ($\sigma_{A} = 0.4$,
$\sigma_{0} = 0.6$, $\sigma_{B} = 0.2$, $\tau = 0.2$) the estimate of
$\sigma_{0}$ is biased upward about 13\%, and when all four PSDs are
large ($\sigma_{A} = 1.0$, $\sigma_{0} = 1.2$, $\sigma_{B} = 0.8$,
$\tau = 0.8$) it is biased downward by about the same percentage. The
direction of the biases implies that estimates tend to be somewhat more
central with respect to the restricted range imposed ($0 < \sigma_{0} <
1.5$) than the true value, thus moderating the estimates. The
coefficient of variation of $\sigma_{0}$ is about 35\% in the former
case and about 19\% in the latter case. This corresponds to roughly the
same standard deviation of the estimate of $\sigma_{0}$ whether
$\sigma_{0}$ is 0.6 or~1.2.

This effect only shows up with respect to $\sigma_{0}$; the other
columns in Display~\ref{tab5} show that mean estimates of $\sigma_{A}$,
$\sigma_{B}$ and $\tau$ are relatively unaffected by any of the
three factors in the simulation,\vadjust{\goodbreak} especial\-ly~$\sigma_{A}$ and
$\sigma_{B}$. Consideration of degrees of freedom may explain
this---these two variance components have $(G - J)(K - 1)$ degrees of
freedom, whereas $\sigma_{0}$ has $K - 1$ df and $\tau$ has $G - J$ df, so
one might expect them to be harder to estimate (although the definition
of degrees of freedom is somewhat fuzzy in this nonlinear Bayesian
setting). Estimates of $\tau$ seem to be most variable percentagewise,
with coefficient of variation in the 30--40 percent range. In all
cases the coefficient of variation is larger for the smaller true PSDs.
The standard deviation of estimation decreases when the true PSD
decreases, but not fully proportionally.

However, remember that the goal of the analysis is not to estimate the
variance components per se, but to use them to define a model that can
better estimate the logistic coefficients by adjusting to global
patterns in the data across responses and predictor categories. Each
individual estimation does not assume that the PSDs are exactly equal to
their posterior mean, but rather the estimation involves an integration
across the posterior distribution of the PSDs. In that respect, it is
interesting to examine the posterior standard deviations of the PSDs.
They have not been included in Display~\ref{tab5} in order to save space, but in
fact the average of the posterior standard deviations across simulations
was remarkably similar to the sampling standard deviations of the
posterior means of each PSD. They typically differed by only 10\% or so
for each of the 8 sets of 250 simulations. Thus, our model expects that
the PSDs will be hard to estimate and works within that
uncertainty.\looseness=-1

\begin{figure*}
\def\arraystretch{0.97}
\tablewidth=\textwidth
\tabcolsep=0pt
  \begin{tabular*}{\textwidth}{@{\extracolsep{4in minus 4in}}lcccccd{2.3}cc@{}}
\hline
\multicolumn{1}{l}{\textbf{(a)}} &
\multicolumn{4}{c}{\textbf{Treatment effect prior mean} $\boldsymbol{B_0}$}&
 \multicolumn{4}{c}{\textbf{Treatment effect for responses} $\boldsymbol{\beta_{0k}}$}
\\
\ccline{2-5,6-9}
 & \textbf{RMSE} & $\boldsymbol{Z^{2}}$ & \textbf{CI.05} & \textbf{CI.95} & \textbf{RMSE} & \multicolumn{1}{c}{$\boldsymbol{Z^{2}}$} & \textbf{CI.05} & \textbf{CI.95}
 \\
 \hline
All RLR simulations & 0.719 & 0.192 & 0.000 & 0.003 & 1.066 & 26.026 & 0.108 & 0.354\\
All MBLR simulations & 0.383 & 1.167 & 0.056 & 0.070 & 0.466 & 1.248 & 0.061 & 0.074\\
Responses: Frequent & 0.424 & 1.235 & 0.068 & 0.067 & 0.314 & 1.200 & 0.059 & 0.073\\
Responses: Rare & 0.343 & 1.099 & 0.044 & 0.073 & 0.619 & 1.297 & 0.063 & 0.075\\
Mean effects: Zero & 0.386 & 1.163 & 0.046 & 0.076 & 0.491 & 1.284 & 0.052 & 0.090\\
Mean effects: Large & 0.381 & 1.170 & 0.066 & 0.064 & 0.442 & 1.212 & 0.070 & 0.058\\
Prior SDs: Small & 0.286 & 1.023 & 0.043 & 0.062 & 0.375 & 1.217 & 0.058 & 0.073\\
Prior SDs: Large & 0.481 & 1.310 & 0.069 & 0.078 & 0.557 & 1.280 & 0.064 & 0.076\\
\hline
\\
\hline
\multicolumn{1}{l}{\textbf{(b)}} &
\multicolumn{4}{c}{\textbf{Covariate effect prior means} $\boldsymbol{A_g}$} &
\multicolumn{4}{c}{\textbf{Covariate effect for responses} $\boldsymbol{\alpha_{gk}}$}
\\
\ccline{2-5,6-9}
 & \textbf{RMSE} & $\boldsymbol{Z^{2}}$ & \textbf{CI.05} & \textbf{CI.95} & \textbf{RMSE} &  \multicolumn{1}{c}{$\boldsymbol{Z^{2}}$} & \textbf{CI.05} & \textbf{CI.95}
 \\
 \hline
All RLR simulations & 0.490 & 0.105 & 0.000 & 0.000 & 0.819 & 11.662 & 0.166 & 0.190\\
All MBLR simulations & 0.297 & 0.972 & 0.044 & 0.052 & 0.373 & 1.041 & 0.049 & 0.057\\
Responses: Frequent & 0.323 & 0.959 & 0.043 & 0.052 & 0.280 & 1.041 & 0.049 & 0.056\\
Responses: Rare & 0.272 & 0.986 & 0.045 & 0.052 & 0.466 & 1.042 & 0.049 & 0.057\\
Mean effects: Zero & 0.300 & 0.959 & 0.042 & 0.051 & 0.388 & 1.026 & 0.047 & 0.056\\
Mean effects: Large & 0.295 & 0.986 & 0.046 & 0.053 & 0.358 & 1.056 & 0.051 & 0.057\\
Prior SDs: Small & 0.205 & 0.990 & 0.044 & 0.055 & 0.283 & 1.048 & 0.050 & 0.057\\
Prior SDs: Large & 0.390 & 0.954 & 0.043 & 0.049 & 0.463 & 1.034 & 0.048 & 0.057\\
\hline
\\
\hline
\multicolumn{1}{l}{\textbf{(c)}} &
\multicolumn{4}{c}{\textbf{Interaction effect prior means} $\boldsymbol{B_g}$} &
\multicolumn{4}{c}{\textbf{Interaction effect for responses} $\boldsymbol{\beta_{gk}}$}
\\
\ccline{2-5,6-9}
 & \textbf{RMSE} & $\boldsymbol{Z^{2}}$ & \textbf{CI.05} & \textbf{CI.95} & \textbf{RMSE} &  \multicolumn{1}{c}{$\boldsymbol{Z^{2}}$} & \textbf{CI.05} & \textbf{CI.95}
 \\
\hline
All MBLR simulations & 0.347 & 3.189 & 0.151 & 0.144 & 0.346 & 1.116 & 0.059 & 0.057\\
Responses: Frequent & 0.353 & 2.940 & 0.144 & 0.141 & 0.292 & 1.110 & 0.059 & 0.056\\
Responses: Rare & 0.340 & 3.439 & 0.159 & 0.147 & 0.399 & 1.122 & 0.059 & 0.057\\
Mean effects: Zero & 0.347 & 3.114 & 0.150 & 0.140 & 0.360 & 1.120 & 0.060 & 0.057\\
Mean effects: Large & 0.346 & 3.265 & 0.152 & 0.148 & 0.331 & 1.112 & 0.059 & 0.056\\
Prior SDs: Small & 0.171 & 2.489 & 0.130 & 0.122 & 0.203 & 1.174 & 0.062 & 0.061\\
Prior SDs: Large & 0.522 & 3.890 & 0.173 & 0.166 & 0.488 & 1.058 & 0.056 & 0.053\\
\hline
\end{tabular*}
\caption{Summary of estimated logistic coefficient distributions
within the simulations. Separate subtables for \textup{(a)}~treatment effects
$B_{0}$ and $\beta_{0k}$, \textup{(b)} covariate main effects $A_{g}$ and
 $\alpha_{gk}$, \textup{(c)} treatment-by-covariate interactions $B_{g}$ and
 $\beta_{gk}$ ($g = 1, \ldots, G$). See text for explanation of the
summary statistics.}\label{tab7}\vspace*{-6pt}
\end{figure*}

\subsection*{Estimation of Logistic Coefficients}

Display~\ref{tab7} summarizes the simulation distributions of the various
logistic regression coefficients. Part (a) of Display~\ref{tab7} focuses on the
main effect of Treatment. The first two rows of Display~\ref{tab7}(a) compare
the Treatment effect accuracy of the RLR estimates to that of the MBLR
estimates. The first four columns refer to the estimation of the prior
mean coefficient, $B_{0}$, what might be called the ``all response
summary,'' while the last four columns refer to the estimation of
coefficients, $\beta_{0k}$, for the individual responses. Across all
2000 simulations, the RMSE for RLR is almost double that of MBLR for
estimation of $B_{0}$, and more than double, on average, for estimating
the $\beta_{0k}$. Since statistical efficiency is typically
inversely proportional to the square of RMSE, this implies that MBLR is
about 4 times as efficient as RLR at estimating treatment/comparator
odds ratios in this setting.\vadjust{\goodbreak}

\begin{figure*}[b]
\def\arraystretch{0.97}
\tablewidth=\textwidth
\tabcolsep=0pt
\begin{tabular*}{\textwidth}{@{\extracolsep{\fill}}lcd{2.3}cccc@{}}
\hline
 & \textbf{True Int.} & \multicolumn{1}{c}{\textbf{Bias}} & \textbf{RMSE} & $\boldsymbol{Z^{2}}$ & \textbf{CI.05} & \textbf{CI.95}\\
\hline
All MBLR simulations & 0.976 & 0.004 & 0.173 & 1.314 & 0.070 & 0.070\\
Responses: Frequent & 0.985 & 0.007 & 0.179 & 1.346 & 0.078 & 0.077\\
Responses: Freq${}+{}$Rare & 0.968 & 0.000 & 0.167 & 1.281 & 0.061 & 0.062\\
Mean effects: Zero & 0.972 & 0.011 & 0.171 & 1.312 & 0.074 & 0.068\\
Mean effects: Large & 0.980 & -0.004 & 0.175 & 1.315 & 0.065 & 0.071\\
Prior SDs: Small & 0.337 & 0.019 & 0.163 & 1.600 & 0.084 & 0.093\\
Prior SDs: Large & 1.616 & -0.011 & 0.184 & 1.027 & 0.055 & 0.046\\
\hline
\end{tabular*}
\caption{Simulation of the resistance to multiple comparisons bias of
MBLR. At each simulation, the most significant treatment $\times$ covariate
interaction was singled out across all responses by selecting the
largest of the GK values ($G = 16$, $K = 5$ or 10, $GK = 80$ or 160) of
(estimated interaction coefficient)${/}$(estimated posterior s.d. of
coefficient). The MBLR estimates are unbiased with relatively small
RMSE. See text for discussion of other columns.}\label{tab6}
\end{figure*}

The statistic $Z^{2}$ is designed to measure the calibration of the
posterior standard deviations computed by a method to the actual
sampling distribution, where $Z^{2} = 1$ implies perfect calibration.
When $Z^{2} \gg 1$, the claimed standard errors of coefficients are too
optimistic (too small), and the reverse is true when $Z^{2} \ll 1$. The
values of $Z^{2}$ provide similar information to the counts of times
confidence intervals fail to enclose the true values of coefficients.
When $Z^{2}$ is too large and putative standard errors are too small,
the too-short confidence intervals will miss the true values more than
the nominal percent of times, and conversely. Looking at the first two
rows of Display~\ref{tab7}(a), we see that RLR is poorly calibrated in this
sense. Computed standard errors are too large for the all-response
summary and too small for the individual response treatment effects. As
a result, supposedly 90\% confidence intervals had 99.7\% coverage for
the all-response summaries and only 53.8\% coverage for individual
response treatment effects. In contrast, the MBLR estimates are much
better calibrated, with $Z^{2}$ about 1.2 and nominally 90\% intervals
having coverage probabilities averaging about~87\%.

The remaining rows of Display~\ref{tab7}(a) show the behavior of the MBLR
estimation for subsets of simulations defined by the three two-level
factors. Rows~3 and 4 compare results for simulations with the 5 more
frequent responses to those for the 5 less frequent responses. In the
latter case, although the runs generated all 10 responses and all 10
were used in the analysis, the results in the row labeled ``Responses:
Rare'' are based only on accuracy statistics for the 5 least frequent
responses, in order to better isolate the estimation ability of MBLR for
rare events. We see that in fact the RMSE, $Z^{2}$, and 90\% interval
coverage probabilities are roughly the same for the rare and frequent
events. (Of course, we assume that a run with \textit{only} the five
rare events would lead to much more variable estimation---it is the
ability of the Bayesian algorithm to detect and measure similarities
between frequent and rare events, and to ``borrow strength''
appropriately, that allows such accuracy.) The next two rows of
Display~\ref{tab7}(a) show that whether the true prior means are 0 or not makes no
difference in the estimation properties. The final two rows of Display
\ref{tab7}(a) show that estimation is significantly more accurate when PSDs are
small than when they are large, which make sense, because small PSDs
imply more commonality across the responses, and thus more opportunity
to borrow strength and increase estimation accuracy. But even with the
larger set of PSDs, MBLR quite outperforms RLR.

Display~\ref{tab7}(b) shows the corresponding results for the estimation of
covariate main effects. For this example, the estimation of $A_{g}$
and $\alpha_{gk}$ seems to be more accurate, using either RLR or
MBLR, on average for the 16 covariate effects (indexed by $g$) than it was
for the single main effect of treatment. However, the advantage of MBLR
over RLR is about the same, both in terms of RMSE and in terms of
standard error calibration as measured by $Z^{2}$ and the coverage
probabilities of nominal 90\% intervals.

Display~\ref{tab7}(c) shows the simulation accuracy of estimation of
covariate--treatment interaction coefficients. Since the RLR model does
not estimate interactions, only MBLR results are presented. Looking
first at the right-hand set of four columns in Display~\ref{tab7}(c) that refer
to estimation of the $\beta_{gk}$, all four accuracy measures seem
to mimic the values in Display~\ref{tab7}(b)---it appears that MBLR can estimate
individual covariate--treatment interactions as accurately as the main
effects of covariates. Now looking at the first four columns of Display
\ref{tab7}(c), where the $B_{g}$ are being estimated, the entire column of
RMSE values are about the same as in Displays~\ref{tab7}(a) and~\ref{tab7}(b), so the
posterior mean estimates of the $B_{g}$ are about as accurate as those
of~$B_{0}$ and of $A_{g}$. However, it seems that the posterior
standard deviations of the $B_{g}$ are too optimistic, since the
values of $Z^{2}$ are about 3 times too large, and the error rate of the
corresponding nominal 90\% intervals is about 30\% instead of 10\%. This
result is puzzling and awaits further investigation.

\subsection*{Bayesian Shrinkage Estimates are Resistant to the Multiple Comparisons
Fallacy}

Display~\ref{tab6} shows the remarkable power of Bayesian shrinkage estimates to
avoid bias even in the presence of post-hoc selection of the most
significant of many estimates.\vadjust{\goodbreak} For each simulation, the task is to find
the most significant treatment $\times$ covariate interaction among all the
responses. There are 16 covariate-based subsets in the model that get
interaction estimates for every response variable and $K = 10$ or 5
responses, making a total of $16K = 80$ or 160 ratios (estimated
interaction coefficient)$/$(estimated s.e. of interaction coefficient).
The largest ratio (one-sided alternative) in each MBLR analysis is
selected and then the known true value for the selected interaction is
used to compute the accuracy measures. This selection and assessment is
repeated for each of the 2000 simulations. The first column of Display
\ref{tab6} is the average of the true coefficients for the selected
interactions. Remembering that the true interactions are generated from
a $N(0, \sigma_{B}^{2} + \tau^{2})$ distribution, where
$\sqrt{(\sigma_{B}^{2} + \tau^{2})} =\mathrm{either}$ 0.283 (smaller PSDs)
or 1.13 (larger PSDs), it is clear from the ``True Int'' column that
MBLR is selecting fairly large interactions. The column headed BIAS
contains the average difference between the selected estimate and its
true value. Remarkably, the MBLR post-hoc selections have virtually no
bias, either overall or in any of the six factor-based subsets. The
final four columns in Display~\ref{tab6} show the same accuracy measures as
those of Display~\ref{tab7}. The RMSE values in Display~\ref{tab6} are smaller than any
of those in Display~\ref{tab7}, which at first might seem surprising, but is a
consequence of the fact that the maximum of 80 or 160 identically
normally distributed variates will have smaller standard deviation than
a single such variate, due to the short tail of the normal distribution.
The calibration of the posterior standard deviations of the selected
most significant interaction, as measured by the values of $Z^{2}$ and
the coverage probabilities, is not perfect but is similar to that of the
treatment main effects in Display~\ref{tab7}(a). This excellent accuracy of MBLR
shrinkage estimates in the face of post-hoc selection is in spite of the
fact that the variance components which determine the amount of
shrinkage were not known in advance but were estimated separately for
each simulation.

\subsection*{Discussion of the Simulation}

It should not be surprising that data generated by a specific Bayesian
model can be better analyzed by fitting that model. But these
simulations show that there is a surprisingly large advantage to doing
so, and that you give up a lot of efficiency (equivalently waste
clinical resources) by forgoing such an analysis if, in fact, such a
model is realistic. With the RLR approach, which itself is probably a
more efficient analysis than\vadjust{\goodbreak} straight logistic regression, you give up
the possibility of estimating treatment--covariate interactions and yet
still lose accuracy in estimation of main effects. The principal
nonBayesian alternative is to form a single pooled response, treating
the different issues as equivalent. But then you don't even get
estimates for the separate issues and you would be submerging completely
the medical distinction between, say, such a serious adverse event as
Anuria and Dry mouth or Thirst. Our methodology is a ``Goldilocks
alternative'' to the bias-variance trade-off, neither as variable as
estimating so many parameters with no prior shrinkage, nor as biased as
assuming that all issues have the same response to treatment and that
all interactions are 0.

\section{Summary and Conclusions}

Safety issues with low observed frequencies will produce standard
logistic regression estimates with wide confidence intervals (based on
highly variable sampling distributions). Clinical safety data is often
of very fine granularity. Each observation of a subject's adverse event
is described with great precision, providing a great multiplicity of
events to be tabulated and whose event frequencies must be compared
across treatment arms. Defining event groupings for the purpose of
getting pooled events with more reliable relative frequencies is hard to
do in advance, before the set of somewhat frequent events is observed.
After the data are collected, it can be controversial to lump events
together because the selection of events to pool can determine how
significant Treatment/Comparator odds ratios become. The multivariate
Bayesian logistic regression methodology described here is designed to
be a compromise between separate analyses of finely distinguished events
and a single analysis of a pooled event. It requires the selection of a
set of medically related issues, potentially exchangeable with respect
to their dependence on treatment and covariates.

A key concept underlying the proposed methodology is that a set of $K$
issues have been prespecified as important and likely to be biologically
and clinically related. It would be a misuse of the method to try very
many subsets of a large set of issues, stopping only when an
``interesting'' result is obtained. A similar caution pertains to
selection of covariates---only those with some prior justification
should be included. When too many extraneous covariates are entered, the
estimated variance components may lead to over-shrinking those effects
and/or interactions that are present, and lead to overly narrow
confidence intervals.\vadjust{\goodbreak}

The methodology is exploratory in nature, in that the analyst is
encouraged to examine the relationship of the adverse event frequencies
to multiple covariates and to treatment by covariate interactions. These
more complicated models may not be estimable by a standard logistic
regression algorithm because the data are often too sparse for the
number of parameters being estimated. Two strategies are used to cope
with this sparsity. First, a Bayesian model allows the analysis of each
issue to borrow strength from the other issues, assumed medically
related so that this sort of averaging is not unreasonable. The fitting
of the MBLR model is accomplished by the multiple runs of a maximum
likelihood algorithm, together with the estimation of Bayes factors for
a range of values of the unknown variance components. The MBLR algorithm
is intended to be able to measure the degree to which the issues have
similar main effects and interactions with treatment on the logit scale.
The hierarchical prior specification in equations (\ref{eq3})--(\ref{eq5}) allows for
partial averaging across issues for those model coefficients that seem
similar. There is also a tendency for the treatment~$\times$\linebreak[2] covariate
interaction coefficients to be shrunk toward the null value of 0, to an
extent controlled by an estimated EB variance parameter as in (\ref{eq6}). This
shrinkage is intended to offset the tendency of exploratory methods to
find ``significant'' subgroup effects purely by chance. Second, a
comparison method, denoted regularized logistic regression, sets
particular values of the variance components in the empirical Bayes
model to emulate standard logistic regression (without interactions)
while avoiding computational problems and inestimable effects that can
be caused by low counts. This modification is designed to hardly affect
the estimates from standard logistic regression when the data are not
sparse.

Since treatment-by-covariate interaction coeffi-\break cients~are difficult to
interpret, the sums of the treatment main effect plus the covariate
interactions are also presented and interpreted as the estimated
treatment effect that would hold for subjects in the subgroup identified
by the covariate value. This combined effect is apropos to the search
for a subject subgroup that might be particularly vulnerable to an
adverse reaction to treatment.

The goal is to allow safety review of a large amount of clinical data
using a sophisticated methodology that can nevertheless be mastered by
those without advanced training\vadjust{\goodbreak} in Bayesian methods or the theory of
variance component estimation, or the interpretation of large masses of
sparse data. Section~\ref{sec5} shows an example partial analysis of 10
medically related issues within a pool of 8 studies involving over 5700
subjects with a model involving treatment, 5 covariates involving 13
defined covariate values, and including examination of
treatment-by-covariate interactions. Section~\ref{sec7} describes a
simulation study that measures the large gains in efficiency that MBLR
can attain, compared to separate analyses for each issue. The striking
results in Display~\ref{tab6} show the ability of Bayesian modeling to greatly
reduce bias due to post-hoc selection of the most significant contrast.

The Multivariate Bayesian Logistic Regression is a technique that can
add to the tools available to the data analyst or medical reviewer. The
method does not eliminate the need for experimental replicability and
convergence with medical knowledge. A~significant Bayesian result found
in one sample that is not replicable may just be indicative of
a~sampling problem. With that said, it is hoped that this new tool will
ease the burden of seeing the forest for the trees during the analysis
of clinical safety data.\looseness=1\vspace*{3pt}

\section*{Acknowledgments}\vspace*{3pt}

Thanks to Sally Cassells, Rick Ferris and Rave Harpaz for their data
management and computer assistance, and to Ram Tiwari and Brad McEvoy
for extensive and helpful comments on an earlier version of this paper.
Any remaining flaws in the conception and execution of this research are
solely due to the author.\vspace*{3pt}


\end{document}